\documentclass[sigconf]{acmart} 

\pdfoutput=1
\usepackage{booktabs} 
\usepackage{amsfonts}       
\usepackage{nicefrac}       
\usepackage{microtype}      
\usepackage{graphicx}      
\usepackage{algorithm}      
\usepackage{algpseudocode}      

\usepackage{listings, lstautogobble, lstlinebgrd}
\usepackage{balance}  
\usepackage{amsmath}
\usepackage{amssymb}
\usepackage{framed}
\usepackage{float}
\usepackage{multirow}
\usepackage{url}
\usepackage{courier}
\usepackage{stfloats}
\usepackage{color}
\usepackage{caption}
\usepackage{subcaption}
\usepackage{cleveref}
\usepackage{diagbox}
\usepackage{tablefootnote}
\usepackage{xspace}
\usepackage{enumitem}

\DeclareMathOperator*{\argmax}{arg\,max} 

\definecolor{codegreen}{rgb}{0,0.6,0}
\definecolor{codegray}{rgb}{0.5,0.5,0.5}
\definecolor{codepurple}{rgb}{0.58,0,0.82}
\definecolor{backcolour}{rgb}{0.92,0.92,0.92}

\lstdefinestyle{mystyle}{
    frame=single,
    commentstyle=\color{codegreen},
    keywordstyle=\color{magenta},
    numberstyle=\tiny\color{codegray},
    stringstyle=\color{codepurple},
    basicstyle=\ttfamily\footnotesize{},
    breakatwhitespace=false,         
    breaklines=true,                 
    captionpos=b,                    
    keepspaces=true,                 
    numbers=left,                    
    showspaces=false,                
    showstringspaces=false,
    showtabs=false,                  
    tabsize=2,
    escapeinside={<@}{@>}
}

\renewcommand\footnotetextcopyrightpermission[1]{} 
\newcommand{\cuttlefish}{Cuttlefish\xspace}

\setcopyright{rightsretained} 

\acmDOI{10.475/123_4} 

\acmISBN{123-4567-24-567/08/06} 

\acmConference[WOODSTOCK'97]{ACM Woodstock conference}{July 1997}{El Paso, Texas USA} 
\acmYear{1997} 
\copyrightyear{2016} 

\acmPrice{15.00} 

\begin{document}
\title{\cuttlefish{}: A Lightweight Primitive for Adaptive Query Processing}


\author{Tomer Kaftan}
\affiliation{%
\institution{University of Washington}
}

\author{Magdalena Balazinska}
\affiliation{%
\institution{University of Washington}
}

 \author{Alvin Cheung}
\affiliation{%
  \institution{University of Washington}
}

 \author{Johannes Gehrke}
\affiliation{%
  \institution{Microsoft}
}

\renewcommand{\shortauthors}{T. Kaftan et al.}





\renewcommand{\shortauthors}{T. Kaftan et al.}

\begin{abstract}
Modern data processing applications execute increasingly sophisticated
analysis that requires operations beyond traditional relational algebra. As a result, operators in query plans grow in diversity and complexity. Designing query optimizer rules and cost models to choose physical operators for all of these novel logical operators is impractical. To address this challenge, we develop \cuttlefish{}, a new primitive for adaptively processing online query plans that explores candidate physical operator instances during query execution and exploits the fastest ones using multi-armed bandit reinforcement learning techniques. We prototype \cuttlefish{} in Apache Spark and adaptively choose operators for image convolution, regular expression matching, and relational joins. Our experiments show \cuttlefish{}-based adaptive convolution and regular expression operators can reach 72-99\% of the throughput of an all-knowing oracle that always selects the optimal algorithm, even when individual physical operators are up to 105$\times$ slower than the optimal. Additionally, \cuttlefish{} achieves join throughput improvements of up to 7.5$\times$ compared with Spark SQL's query optimizer.

\end{abstract}

%
%

\maketitle

\section{Introduction}

Modern data processing applications execute many operators from a variety of domains.  In addition to standard relational operators~\cite{armbrust2015spark}, these applications utilize operators for scientific analysis~\cite{mehta2016comparative, nothaft2015rethinking}, graph processing~\cite{gonzalez2014graphx, Murray:2013:NTD:2517349.2522738}, and machine learning~\cite{meng2016mllib, ghoting2011systemml} among others. Existing general-purpose, distributed, shared-nothing frameworks with flexible programming models~\cite{zaharia2016apache, carbone2015apache, Akidau:2015:DMP:2824032.2824076, rocklin2015dask} facilitate the implementation of novel, user-defined operators. A key challenge
is that application developers must choose between multiple algorithms and implementations for each of these operators. However, the fastest implementation may depend on the hardware, workloads, and underlying input data. To make matters worse, these properties may change over time or vary across machines in a distributed cluster.

Traditional database management systems (DBMSes) manage to automatically pick operator implementations for users, by allowing users to write queries using high-level languages, then using rule-based and cost-based query optimizers~\cite{chaudhuri1998overview} to generate efficient physical query plans. Unfortunately, designing effective rules and cost-based models for query optimizers is a challenging, time-consuming process. It is crucial to get this process right, because poorly generated query plans can drastically harm performance~\cite{leis2015good}. The query optimizer design process is so labor-intensive that Spark SQL~\cite{armbrust2015spark} lacked cost-based query optimization for two years after its initial release.\footnote{http://databricks.com/blog/2017/08/31/cost-based-optimizer-in-apache-spark-2-2.html} While it is feasible to design effective query optimizers for the small set of relational operators, it is impractical for system developers to design effective rules and models as the operators in modern applications grow in number and in complexity. Additionally, traditional query optimizers do not adapt to changing workloads; they select a single physical query plan and process every record using the same operator implementations. Adaptive query processing (AQP) techniques aim to modify query execution at runtime~\cite{deshpande2007adaptive, babu2005adaptive, hellerstein2000adaptive} in response to optimizer errors and to adapt to dynamically changing workloads. Unfortunately, just like traditional query optimizers, almost all existing AQP approaches still require system developers to have deep knowledge of the underlying operators and to design explicit rules and cost models.

Recently, prior work in Vectorwise introduced the idea of \textit{Micro Adaptivity}, an AQP approach that automatically chooses between several implementations of operators during query execution time, without requiring any explicit optimization rules~\cite{ruaducanu2013micro}. Instead, Micro Adaptivity uses \textit{Multi-armed bandit} (MAB) solvers to tune the choice of operator implementations online. MAB solvers use adaptive learning techniques related to reinforcement learning~\cite{burtini2015survey,auer2002finite,bubeck2012regret} that work by balancing the exploration and exploitation of different alternatives. Yet, the initial approach posed in Vectorwise has several limitations. 
To begin, it is limited to execution models that use blocks of uninterleaved operator execution, such as vectorized execution models. It uses MAB algorithms whose performance is highly sensitive to how meta-parameters governing the amount of exploration are configured. It is also not designed for distributed shared-nothing settings. Finally, it does not support incorporating knowledge about operators and data properties when it is available. We build on the idea of Micro Adaptivity, but provide an approach that vastly expands its applicability.

Designing an effective approach for tuning operators online comes with many challenges. First, the core tuning approach must be flexible enough to work for as many operators and execution models as possible, e.g., fine-grained operators that process data tuple-at-a-time, coarser-grained vectorized operators, stateful operators, distributed parallel operators that involve data shuffles, pipelined operators, out-of-core operators, etc. The tuning algorithm itself should be computationally lightweight, and it should identify the fastest operator implementations quickly and accurately, without requiring developers to tweak the algorithm parameters. If developers had to carefully refine the parameters for every single operator, it would counteract the benefits of having a single primitive that can tune any operator.

When no input data statistics are available, the tuning approach should identify the fastest-on-average operator implementation for the workload. Where possible, the tuning algorithm should automatically learn cost models specifying how the runtimes for each operator implementation vary according to the input data properties, and use these models to drive physical operator selection. The approach needs to tune effectively as cluster sizes grow in distributed shared-nothing environments, without introducing excessive system overheads or impacting workload scalability. Finally, the tuning approach needs to be able to adapt to dynamic workloads that change over time, or vary across machines and cores in the cluster.

In this paper we present \cuttlefish{}, a new primitive for adaptive query processing. We focus on query plans with already-chosen operator orderings and demonstrate how developers can use \cuttlefish{} to tune operators that process data using a variety of granularities and execution models. Developers can call \cuttlefish{}'s simple yet flexible API to insert online tuners anywhere in their applications. As the operators execute, the \cuttlefish{} tuners choose and execute one of several candidate physical operator fragments on subsets of the input data at granularities supported by the physical operators. For example, a \cuttlefish{} tuner can pick one algorithm per image for a convolution operator, and one join strategy per data partition for a distributed parallel join operator. The tuners balance exploration and exploitation to quickly identify the fastest physical operators.

\cuttlefish{} tuners rely on a class of probability-matching MAB techniques known as \textit{Thompson sampling}. These algorithms are theoretically-sound~\cite{agrawal2012analysis,agrawal2013further,kaufmann2012thompson,korda2013thompson} and empirically effective~\cite{scott2010modern,chapelle2011empirical}. They allow \cuttlefish{} to tune operators with drastically different runtime characteristics, without requiring any manual parameter tweaking by developers. When application developers can collect relevant context about the input data and the workload, \cuttlefish{}'s tuners learn to contextually select the best physical operator for each part of the workload rather than just using the single best-on-average physical operator. \cuttlefish{} does this by learning cost models entirely online that capture how operator implementations' runtimes depend on the context. \cuttlefish{} tuners can then use these cost models to pick the best physical implementation according to the context

\cuttlefish{} uses a distributed tuning architecture that allows tuners to learn from operator execution across the entire cluster instead of just one core, without requiring excessive communication nor imposing synchronization overheads. Additionally, \cuttlefish{} tuners can adapt to dynamically changing workloads and to workloads where operators executing on different machines in the cluster have different runtime characteristics.

We prototype \cuttlefish{} in Apache Spark~\cite{zaharia2016apache} and tune operators for image convolution, regular expression matching, and relational joins. On real datasets and the TPC-DS benchmark we observe significant speedups for all three operators.

To summarize, this paper makes the following contributions:
\begin{itemize}[leftmargin=0cm]
  \item We present \cuttlefish{}, a primitive that enables adaptive query processing without explicit optimization rules or cost models. We demonstrate how to use \cuttlefish{}'s concise, powerful API to tune operators with different granularities and execution models. (\Cref{sec:cuttlefishapi})
  \item We provide computationally-lightweight Thompson sampling ``multi-armed bandit'' algorithms for tuning. These are backed by useful theoretical guarantees, and they effectively tune drastically different operators without any per-operator algorithm tweaking. These algorithms can even learn and use cost models for contextual tuning automatically, entirely online. (\Cref{sec:tuningpolicies})
  \item We present a distributed tuning architecture for \cuttlefish{} that learns from workload execution throughout the entire shared-nothing cluster, without incurring excessive system overheads. (\Cref{sec:sharednothing})
  \item We describe how \cuttlefish{} tuners can adapt to workloads that dynamically change over time, and vary across machines or cores in a cluster. (\Cref{sec:nonstationaryrewards})
  \item We prototype \cuttlefish{} in Apache Spark, and tune operators that have several implementations.
  Our experiments show join throughput improvements of up to 7.5$\times$ compared with Spark SQL's query optimizer. Additionally, convolution and regular expression operators tuned with \cuttlefish{} can reach 72-99\% of the throughput of an all-knowing oracle that always selects the optimal algorithm, even when individual convolution operators are up to 6$\times$ slower than the optimal and some regular expression operators are up to 105$\times$ slower than the optimal.
  We also explore how the effectiveness of online tuning varies for different operators and workloads. (\Cref{sec:evaluation})
\end{itemize}

\section{Motivating Example}
\label{sec:motivation}

\begin{figure}[!t]
  \centering
  \includegraphics[width=1.0\columnwidth]{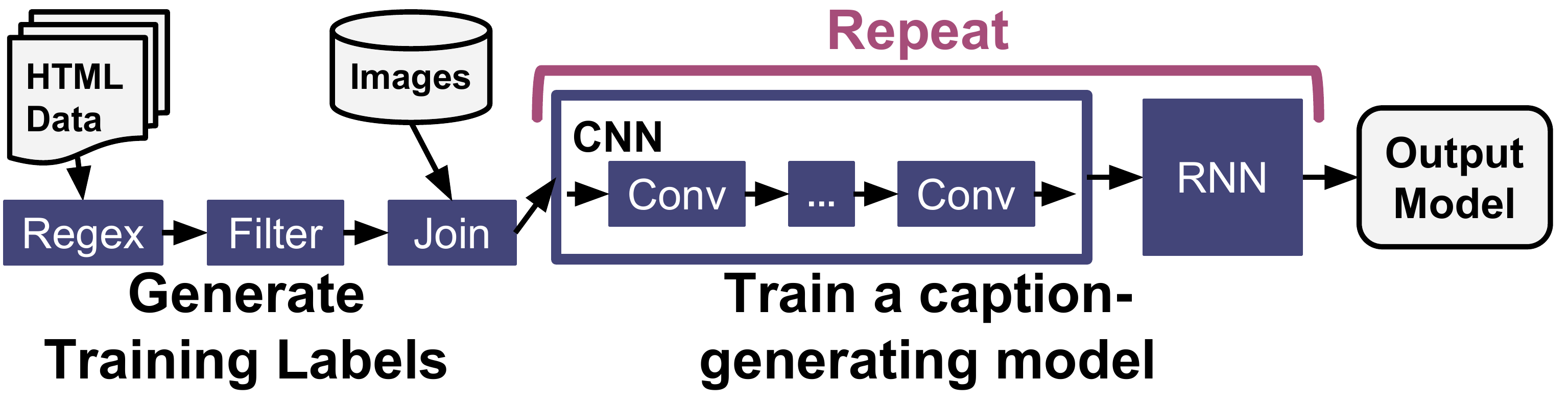}
  \vspace{-0.2in}
\caption{Example query plan for application that generates captions for images on the Web}
\label{fig:motivatingexample}
\vspace{-0.15in}
\end{figure}

\begin{figure}[!t]
  \centering
  \includegraphics[width=1.0\columnwidth]{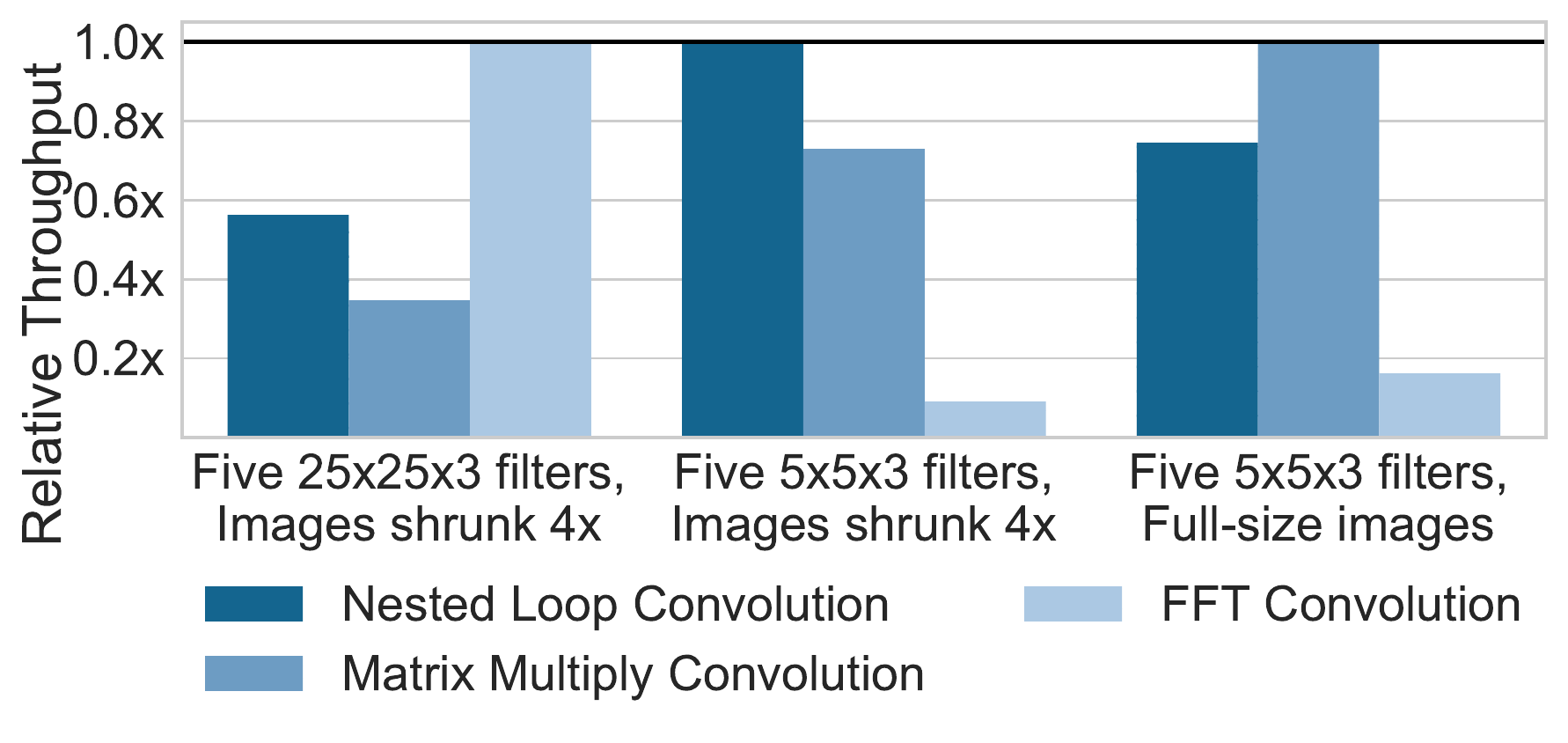}
  \vspace{-0.25in}
\caption{Relative throughput of three different convolution algorithms within different workloads. 1x represents the throughput of the fastest tested algorithm for a workload.}
\label{fig:flickr_8node_motivation}
\vspace{-0.15in}
\end{figure}

Modern data processing applications use a diverse set of operators from many different domains. As an example, consider a data scientist who needs to train a neural network for generating comments on images.  She first uses regular expressions (regex) to extract image comments from HTML documents (obtained from a Web crawl) to serve as training labels~\cite{chang2006survey}. She filters out poorly rated comments and generates the training pairs by joining the extracted comments with the images themselves.
The data scientist then trains a caption-generating model~\cite{xu2015show} comprising a convolutional neural network (CNN) that learns to extract image features, with a recurrent neural network (RNN) that learns to generate text descriptions from those features. This network consists of multiple layers, each of which is a separate operator in the query plan. Many of these layers apply convolutions~\cite{szegedy2016rethinking,krizhevsky2012imagenet}, compute-intensive mathematical operations that apply one or more kernels, i.e., filters, to an input signal. \Cref{fig:motivatingexample} shows a possible query plan for this application.

The above query plan consists of many logical operators: regex matching, relational selects and joins, and convolutions. These logical operators have multiple physical implementations with different performance profiles. Importantly, the performance of these operators often depends on the input parameters and data, making it difficult for users to select implementations. To illustrate this challenge, \Cref{fig:flickr_8node_motivation} shows that the relative throughput of various convolution algorithms varies drastically when convolving 8000 images from Flickr~\cite{hodosh2013framing} using different image scales and sets of filters. For example, an FFT-based convolution~\cite{mathieu2013fast} is fastest when convolving shrunk images with $25\times25\times3$ pixel filters, but if the filters are changed to $5\times5\times3$ pixels then the FFT convolution becomes 10$\times$ slower than a naive nested-loops convolution.

Because this query plan contains diverse, non-relational operators, the query optimizer is unlikely to have the heuristics it needs to select a good physical plan. The data scientist should not have to design cost models and optimization rules herself for every logical operator in her workload that the query optimizer does not have pre-existing heuristics for. It would also be wasteful for her to tune this workload offline~\cite{ansel2014opentuner} because compute resources spent executing representative workloads to collect profiling information would not go towards training the caption-generating model. Instead, \cuttlefish{}'s flexible API allows the data scientist to tune her workload as it trains the caption-generating model.

\section{\cuttlefish{} API}
\label{sec:cuttlefishapi}

\begin{figure*}[!t]
  \centering
  \includegraphics[width=2.0\columnwidth]{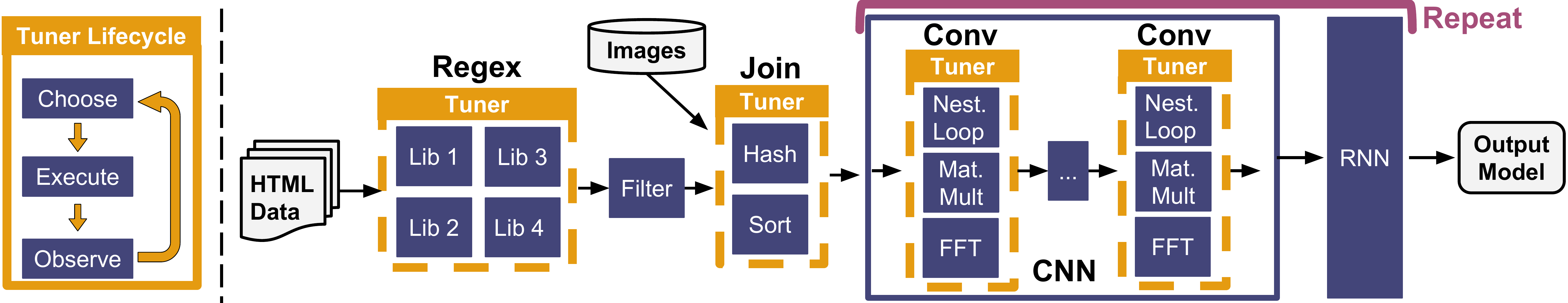}
\caption{Left: The tuner lifecycle. Right: \cuttlefish{} tuning the workload from \Cref{sec:motivation}.}
\label{fig:tuningoverview}
\vspace{-0.15in}
\end{figure*}

\begin{figure}
\begin{lstlisting}[language=python]
<@\label{line:construct}@>class Tuner(choices):
    <@\label{line:choose}@>def choose(context=None) -> (Choice, Token) 
    <@\label{line:observe}@>def observe(token, reward) -> None 
\end{lstlisting}
\vspace{-0.1in}
\caption{The \cuttlefish{} API}
\label{fig:cuttleinterface}
\vspace{-0.15in}
\end{figure}

\cuttlefish{} is a dynamic tuning framework that automatically adapts a physical query plan based on workload characteristics. The \cuttlefish{} API extends the query execution layer of a database system with \textit{tuner} instances that application developers (or the query optimizer) can insert into their query plans. Each tuner is in charge of tuning between a different set of candidate physical operators during query execution. \Cref{fig:tuningoverview} (Left) shows the lifecycle of a \cuttlefish tuner. At every tuning round during query execution, tuners {\it choose} a physical operator, the system {\it executes} this operator, and the tuner {\it observes} a reward for the chosen operator that corresponds to the optimization metric. For example, if the goal is to optimize throughput, physical variants with high throughput will produce large rewards and vice versa for low throughput variants. The tuners use reinforcement learning to try different physical variants and quickly settle on the ``best'' variant, i.e., the operator variant with the biggest expected reward.

Rather than just learning to pick a best-on-average variant, \cuttlefish{} tuners can optionally observe a vector of context features at the start of each tuning round and learn to pick the best variant given the context. For example, if there is one tuning round per tuple and the context vector contains the size of the tuple, the tuner will learn which variant works best for each tuple size.

Specifically, each tuner solves a separate \textit{stochastic multi-armed bandit problem} (MAB problem)~\cite{burtini2015survey,auer2002finite,bubeck2012regret}. In the MAB problem, an agent is provided with a set of arms. Each arm has an initially unknown reward distribution. For every round in a sequence of rounds, the agent must select an arm. At the end of each round, the agent observes a reward sampled from the picked arm's reward distribution. The agent's goal is to maximize the expected total reward across all rounds. The agent does this by balancing the \textit{exploration} of insufficiently explored arms with the \textit{exploitation} of arms that the agent believes have high rewards. For \cuttlefish{} tuners, each candidate physical operator maps to a separate arm in the MAB problem. Tuning with context maps to the contextual MAB problem~\cite{burtini2015survey,li2010contextual,agrawal2013thompson}.

The application developer (or the query optimizer) maps query execution to tuning rounds by inserting calls to \cuttlefish{}'s API into the physical query plan. The API is shown in \Cref{fig:cuttleinterface}. To use the API, a developer (or the query optimizer) first constructs a \texttt{Tuner} instance and provides a set of choices as input. The \texttt{choices} are variables that can point to implementations of a logical operator, but they can be of any type including primitive types such as integers. For example, if a vectorized operator has a batch size parameter, a \cuttlefish \texttt{Tuner} can choose among different candidate batch sizes.

The \cuttlefish{} API separates operator selection from reward observation to give \cuttlefish{} users flexibility in tuning operators with different execution models at various granularities and in choosing which metric to tune for. It also allows users to interleave tuning rounds for different tuners or to interleave rounds for the same tuner across different threads in a cluster. At the start of each tuning round, the developer has the tuner pick one of the choices. The developer does this by inserting calls to \texttt{choose} at the appropriate places in their code, optionally providing a context for each round. At the end of the round, the developer provides a reward to the tuner using \texttt{observe}. The developer might compute this reward from the operator throughput, the operator latency, or any other metric. We recommend setting the reward to be the runtime of the operator during the round multiplied by $-1$. Maximizing the sum negative runtime across all rounds becomes directly equivalent to maximizing the operator throughput.

The \texttt{tokens} in the method definitions are an API convenience that contain the choice, the context, and all other information about a decision that tuners require when observing a reward in \texttt{observe}. These tokens simplify the \cuttlefish API and free \cuttlefish users from manual bookkeeping that depends on the underlying learning algorithm.

\Cref{fig:tuningoverview} (Right) illustrates how the data scientist from \Cref{sec:motivation} might use \cuttlefish{} to tune her workload. As discussed earlier, regular expression matching, relational joins, and convolution all have multiple physical implementations. So, the data scientist (or a query optimizer) assigns a \cuttlefish{} \texttt{Tuner} for each such logical operator in her workload to select between candidate implementations. She then splits the execution of each operator into \cuttlefish{} tuning rounds according to the execution granularities of the operators. For example, she can have the regex tuner select a regular expression library once per document. A distributed parallel join is a stateful operator and may only be able to switch implementations once per data partition; so, the data scientist can map each join partition to a separate tuning round. Finally, she can have a separate tuner select a convolution algorithm once per image for each logical convolution operator in her query plan. Alternatively, if the data scientist has enough context about each image and filter to fully model the convolution runtime of each algorithm, she could have just one contextual convolution \texttt{Tuner} tune all logical convolution operators in her workload at once.

In the following subsections we show in depth how to use \cuttlefish{} to adaptively tune both the convolution operators and the distributed parallel join operator, despite their drastically different granularities and execution models.

\subsection{Basic Operator Tuning}
\label{sec:usingcuttlefish}
\begin{figure}[t]
\centering
\begin{subfigure}{\columnwidth}
\begin{lstlisting}[language=python, escapeinside=||]
# Define the convolution variants
def loop_convolve(image, kernel) -> Image
def mm_convolve(image, kernel) -> Image
def fft_convolve(image, kernel) -> Image

# Define a func. to extract image and filter dims
def extract_dimensions(image, kernel) -> Vector
\end{lstlisting}
\vspace{-0.1in}
\caption{Convolution variants and a function to extract dimensions}
\label{fig:basicuser:define}
\end{subfigure}
\vspace{0.05in}

\begin{subfigure}{\columnwidth}
\begin{lstlisting}[language=python, escapeinside=||, firstnumber=last]
conv_choices = [loop_convolve, mm_convolve, fft_convolve]
tuner = Tuner(choices = conv_choices)

results = []

# images: the list of images to convolve
# kernel: the convolution kernel to use
for image in images:
    dims = extract_dimensions(image, kernel)
    convolve, token = tuner.choose(context = dims)
    
    start = datetime.now()
    convolved_image = convolve(image, kernel)
    end = datetime.now()
    
    results.append(convolved_image)
    tuner.observe(token, reward = start - end)
\end{lstlisting}
\caption{\cuttlefish{} tuner to adaptively convolve the images}
\label{fig:basicuser:withcuttle}
\end{subfigure}

\vspace{-0.1in}
\caption{Adaptively tuning an image convolution operator}
\label{fig:basicuser}
\vspace{-0.1in}
\end{figure}

In this subsection, we show how \cuttlefish{}'s API can serve to adaptively tune convolution operators to optimize for the total operator throughput. There are many known ways to convolve images with kernels, including nested loops, vectorized matrix multiplication~\cite{hadjis2015caffe}, and using the fast Fourier transform (FFT) to convolve in the frequency domain~\cite{mathieu2013fast}. The performance of the convolution variants depends on the image and kernel dimensions.

In \Cref{fig:basicuser:define}, we define functions that convolve images using each of the above three algorithms (Lines~2-4), and to extract the dimensions of images and kernels (Line~9). With these functions defined, \Cref{fig:basicuser:withcuttle} shows how to tune a convolution operator using \cuttlefish. The operator first constructs a \cuttlefish{} tuner on Line~10, using our three algorithm variants as the choices. For each image, the operator explicitly computes the image and kernel dimensions (Line~16), and passes them to the tuner's \texttt{choose} method as context features when picking one of the convolution variants (Line~17). Note that the context computation step is entirely optional, and that the operator can instead call the tuner's \texttt{choose} method with no context provided. In that case, the tuner learns to always select the best-on-average physical implementation. Once the tuner chooses a convolution variant, the operator convolves the image using this variant (Line~20) and adds the output to the results (Line~23). Finally, the operator provides the negative convolution runtime to the tuner as a reward for that round (Lines~19,21,24). This causes the tuner to maximize the total convolution throughput.
\subsection{Complex Operator Tuning}
\label{sec:tuningpipelined}

\cuttlefish can also be used in more complex scenarios where each operator invocation does not fully capture all the work required to process the input. For example, a distributed parallel join operator in pipelined query plans will return an iterator of join results per parallel result partition. Each invocation of a local single-threaded join in that parallel join operator only completes some of the join computation. The rest of the computation occurs as the local result iterators are processed by downstream operators in the query plan. We show how to use \cuttlefish in these scenarios by adaptively tuning a distributed join operator in \Cref{fig:lessbasicuser}.

\begin{figure}[t]
\centering
\begin{lstlisting}[language=python]
def hash_join(expr, left, right) -> Iter[Row]
def sort_merge_join(expr, left, right) -> Iter[Row]

tuner = Tuner(variants = [hash_join, sort_merge_join])

def parallel_join(expr, left_input, right_input):
    partitions = co-partition(left_input, right_input)
    join_result = []
    for (left, right) in partitions:
        local_join, token = tuner.choose()
        start = datetime.now()
        local_result = local_join(expr, left, right)
        
        # Define callback to observe reward later
        def observe_join_time():
            reward = datetime.now() - start
            tuner.observe(token, reward = reward)
        
        # Register callback
        local_result.on_iter_finish(observe_join_time)
        join_result.append(local_res)
        
    return join_result
\end{lstlisting}
\vspace{-0.15in}
\caption{Adaptively tuning a stateful, pipelined operator}
\label{fig:lessbasicuser}
\vspace{-0.15in}
\end{figure}

In this implementation of a distributed join, the operator first partitions the inputs on their join attribute (Line 7) and then selects either hash or sort merge join (Line 10) to apply to each local partition. Depending on the chosen join method, the initial call to \texttt{local\_join} may perform only part of the work (e.g., build the hash table for the build relation or sort the inputs) and return an iterator. 
The reward thus needs to capture the total elapsed time starting when the local join begins processing a local partition and finishing when later operators have fully consumed the iterator of join results for that partition. This is shown in Line 11 and in Lines 14-18 using a callback function that executes when the iterator is consumed. This elapsed time captures the total join execution time. The operator then negates this elapsed time and provides it to the tuner as a reward. This enables the tuner to maximize the join throughput as before. If there is no iterator utility to register callback functions, downstream operators in the plan can directly call \texttt{observe} when they finish consuming iterators of local join results.

\section{\cuttlefish{} Tuning Algorithms}
\label{sec:tuningpolicies}

We now describe the multi-armed bandit learning algorithms that \cuttlefish{} tuners use to select among different physical implementations.
These algorithms aim to allow tuners
to quickly and accurately identify which operator variant is optimal, i.e. has the highest expected reward, even in environments with drastically different workloads, operator variants, and tuning granularities. If developers are tuning with context features, \cuttlefish{} tuners need to learn and utilize cost models to contextually select operator variants. Additionally, because tuning may occur at fine granularities, tuners might pick variants and observe rewards many times within a workload. So, the computational overheads of tuners should be as lightweight as possible. 
Finally, the learned state should be updated in a fully online fashion, and the tuner's computation and memory overheads should not grow over time as more rewards are observed. 

We pose tuning in \cuttlefish{} as a \textit{stochastic multi-armed bandit problem} (MAB problem)~\cite{burtini2015survey,auer2002finite,bubeck2012regret}. We solve the MAB problem with Thompson sampling~\cite{scott2010modern,chapelle2011empirical} in both contextual and non-contextual settings.
For simplicity, in this section we describe the tuning algorithms in single-threaded environments and assume that workloads do not change over time.

\subsection{Bandits \& Thompson Sampling Background}
\label{sec:bandittuning}

The goal of agents that solve the MAB problem is to maximize the total reward after a sequence of independent decision rounds, where each round has a set of choices, called \textit{arms}. Each arm has an initially unknown reward distribution, and rewards are sampled from the distributions of the chosen arms. Algorithms for solving the MAB problem work by balancing \textit{exploration} and \textit{exploitation} in each round. Agents explore by selecting arms to learn more about their reward distributions, and exploit by picking arms they believe have high rewards. \cuttlefish{} supports a variety of bandit heuristics such as \textit{$\epsilon$-greedy} and the \textit{upper confidence bound} (UCB) family of algorithms~\cite{auer2002finite}. We focus on \textit{Thompson sampling}, which has state-of-the-art performance
in practice~\cite{scott2010modern,chapelle2011empirical} and is backed up by theoretical guarantees~\cite{agrawal2012analysis,agrawal2013further,kaufmann2012thompson,korda2013thompson}. Agents that use Thompson sampling randomly choose arms according to the likelihood that they have the highest expected reward, given their initial prior assumptions and the observed rewards.

We can design tuners using Thompson sampling as follows: we denote reward values with $r$ and assume the unknown reward distributions belong to a family of distributions $P(r|\theta)$ which are fully specified by a set of model parameters $\theta$. We also begin with prior assumptions about how likely different values of $\theta$ are. At each round $t$ we construct $P(\theta^*|a,r_0,\dots,r_{t-1})$, a posterior distribution over the parameters of each arm $a$ given the observed rewards $r_0,\dots,r_{t-1}$, and the priors over $\theta$.
We sample candidate model parameters $\hat{\theta}_a$ for each arm. We then pick the arm with the highest expected reward given these parameters, i.e., $\argmax_a\mathbb{E}[P(r_t|\hat{\theta}_a)]$, and continue to the next round. We can use these tuners even when the true reward distributions do not match the initial assumptions. However, Thompson sampling algorithms tune faster and are more accurately when the initial assumptions are realistic~\cite{liu2016prior}.

\subsection{Tuning with Thompson Sampling}
\label{sec:banditnocontext}

\begin{figure}[t]
\centering
\begin{lstlisting}[language=python, mathescape]
class ThompsonSamplingTuner(choices):
    numChoices = len(choices)
    sampleCount = [0 for i in range(numChoices)]
    sampleMean = [0.0 for i in range(numChoices)]
    sampleVar = [0.0 for i in range(numChoices)]

    def choose():
        $\theta$ = {}
        for i in range(len(choices)):
            if sampleCount[i] >= 2:
                $\theta$[i] = t_distribution(
                    $\nu$ = sampleCount[i] - 1,
                    $\mu$ = sampleMean[i],
                    $\sigma^2$ = sampleVar[i] / sampleCount[i]
                ).sample()
            else:
                $\theta$[i] = uniform($-\infty$, $\infty$).sample()
        choice = $\arg\max_i\theta$[i]
        return (choices[choice], {'choice': choice})

    def observe(token, reward):
        choice = token['choice']
        sampleCount[choice] += 1
        update sampleMean[choice] using reward
        update sampleVar[choice] using reward
\end{lstlisting}
\vspace{-0.1in}
\caption{Thompson sampling Tuner without Context}
\label{fig:thompsonsampling}
\vspace{-0.2in}
\end{figure}

To apply Thompson sampling to our setting when there are no context features, we model rewards as Gaussian distributions with initially unknown means and variances~\cite{honda2014optimality}. This is an effective assumption when tuning operators because the runtime distributions of physical operator variants are likely to be peaked, unbounded, and continuous, with an unknown center and spread. It is entirely hyperparameter free because it assumes a \textit{noninformative prior}. That is, before observing any rewards, it does not assume any mean or variance is likelier than another. We do this because both the center and spread of operator runtimes can vary by many orders of magnitude depending on the nature of the operator, the granularity of tuning, and even the units used to measure runtime (e.g., minutes, seconds, or CPU cycles). When we later evaluate this learning algorithm in \Cref{sec:evaluation}, we demonstrate that this hyperparameter-free tuner can tune diverse operators across many settings.

We develop a \cuttlefish{} tuner that uses a constant-memory, fully-online implementation of this algorithm that only needs to process each observation once. \Cref{fig:thompsonsampling} shows the pseudocode. The tuner keeps track of the sample count, sample mean, and unbiased sample variance of the rewards it observes for each operator variant (i.e., each arm). Due to the initial set of guassian assumptions with a noninformative prior, the posterior distributions over the population means of the rewards are t-distributions that can be fully specified using the sample counts, means, and variances~\cite{yang1996catalog}. For physical variants whose rewards have been observed less than twice, the t-distributions become ill-defined and the posterior distribution is instead a uniform distribution over all real numbers.
The tuner initializes the sample statistics in Lines 2 to 5. When the tuner \texttt{observe}s a reward, it updates these variables in a single-pass fashion~\cite{pebay2008formulas} in Lines 23 to 27.
To \texttt{choose} an operator variant, the tuner first samples a candidate mean from every variant's posterior distribution in Lines 10 to 19. It then selects and returns the variant with the largest sampled mean in Lines 20 and 21.

\vspace{-0.1in}
\subsection{Contextual Tuning w/ Thompson Sampling}
\label{sec:contextualtuning}

The context-free tuners that do not use context features can only learn a single best-on-average physical operator.
To improve performance, we extend
\cuttlefish{} to select the best physical operator at each tuning round according to the rounds' context features. For example, a convolution tuner can select the fastest convolution algorithm for every image based on the image and kernel dimensions. We show in \Cref{sec:evalcontext} that \cuttlefish{} can effectively utilize user-provided context features when tuning.

\cuttlefish{}'s contextual tuners automatically learn cost models that predict the runtime of every physical operator variant given the context features \textit{during query execution}. The tuners use these models to select the best operator variants given the current context features. The contextual tuners use contextual MAB algorithms that integrate this model training with the exploration and exploitation of physical operator variants.

Learning complex models requires iterative feature engineering, hyper-parameter tuning, and large training datasets~\cite{domingos2012few}. \cuttlefish{} instead learns simple linear models, which can be learned in an online fashion with limited exploration. More specifically, \cuttlefish{}'s default contextual tuner uses \textit{Thompson sampling with linear payoffs}~\cite{agrawal2013thompson}, a contextual Thompson sampling heuristic that learns regularized linear models. We combine this heuristic with standard feature processing steps, and describe this tuner in further detail in Appendix A. As a result, \cuttlefish{}'s contextual tuners are computationally lightweight, tune effectively even with limited numbers of tuning rounds, and do not require large amounts of feature engineering.

\section{Distributed Tuning}
\label{sec:sharednothing}

Modern data analysis applications execute in shared-nothing database
management systems. Online tuning in distributed settings raises
several additional challenges. First, if a tuner handles each execution thread independently with only thread-local learning and no communication, it will only use a subset of all available information. This will prolong how long tuners must explore before they converge to fast operator variants. This impact will worsen as the size of the cluster grows.

Moreover, communication and synchronization overheads arise when making tuning decisions and sharing feedback across threads in a cluster. These overheads can negatively impact performance. For example, it would be prohibitive for fine-grained operators to communicate with a single centralized tuner after each round, as this would require blocking network calls after every single tuning round on every thread.

Finally, if a tuner makes decisions for multiple threads in parallel, it must make several decisions before getting to observe any rewards. 
This is known as \textit{feedback delay}. A variety of theoretical~\cite{joulani2013online} and empirical~\cite{chapelle2011empirical} results show that MAB algorithms can still solve the MAB problem even with delayed reward feedback, but algorithms with large delays take longer to accurately identify the best arms. So, we need to shorten this delay as much as possible.

\begin{figure}[!t]
  \centering
  \includegraphics[width=1.0\columnwidth]{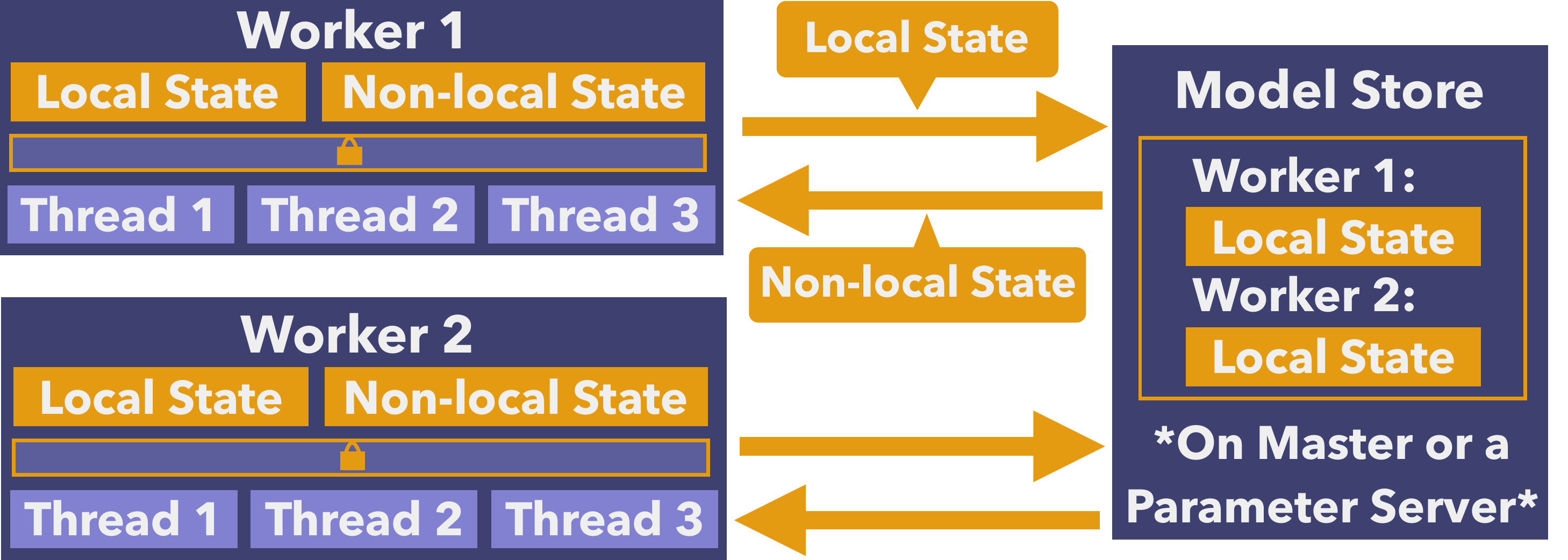}
  \vspace{-0.1in}
\caption{Distributed Tuning in \cuttlefish{}}
\label{fig:sharingfigure}
\vspace{-0.2in}
\end{figure}

To address the first challenge, \cuttlefish{} tuners need to learn from reward observations across the cluster. \cuttlefish{} could do this via three main architectures. (1) \textit{Centralized Tuner}: \cuttlefish{} can direct all \texttt{choose} and \texttt{observe} calls across the network to a single centralized tuner instance. (2) \textit{Thread-local Tuners, Peer-to-Peer Communication}: \cuttlefish{} can assign separate tuner instances to each thread that locally handle calls to \texttt{choose} and \texttt{observe}, and then it can use peer-to-peer algorithms to exchange information between tuner instances. (3) \textit{Thread-local Tuners, Centralized Model Store}: \cuttlefish{} can assign separate tuner instances to each thread and push and pull tuner state updates to and from a centralized model store that aggregates the updates (like a parameter server~\cite{li14scaling}). As we have already mentioned, a single centralized tuner has prohibitively high overheads. Gossip protocols for peer-to-peer bandit learning have been explored~\cite{pmlr-v28-szorenyi13,Korda:2016:DCL:3045390.3045528}, but these have more communication rounds and feedback delays when exchanging information across the cluster than centralized approaches. So, \cuttlefish{} uses independent thread-local tuner instances with a centralized model store, which can reside on
the master node of the underlying distributed execution engine or
on a dedicated parameter server.

To minimize communication and synchronization overheads, \cuttlefish{} tuners at each thread communicate with the centralized model store asynchronously using non-blocking messages, and they only push and pull tuning model state updates at regular intervals rather than every single MAB round. However, to minimize feedback delays, the tuners at each thread should not have to wait until they pull the aggregate global tuning state from the centralized store before they can use their recent local observations when tuning. So, \cuttlefish{} tuners on each thread separately maintain both a locally learned state and a non-local state pulled from the central model store. Tuners merge these two states to make decisions at each MAB round. The local state is immediately updated after every single MAB reward observation. When threads communicate with the central store, they push their local state and pull an aggregated non-local state containing what all other threads have learned. This allows tuners at each thread to learn from their local observations without having to wait for a communication round with the central store. To further minimize the feedback delay, \cuttlefish{} also shares local and non-local tuning states across threads on the same machine, so threads on each machine immediately share what they learn.

\Cref{fig:sharingfigure} shows the overall distributed tuning architecture.
The shared-nothing system comprises a set of workers spread across machines in a cluster. Each worker is a multi-threaded process and each tuner instance executes on one of those threads. Multiple workers can execute on the same machine, but
\cuttlefish{} treats them as if they were executing on different machines.

Within each worker process, threads share a local \texttt{State} that tuners learn only from rewards observed on that worker and a non-local \texttt{State} that represents only observations made on other workers. Sharing \texttt{State} objects between threads on a single worker further minimizes feedback delay between threads. A centralized model store maintains a registry of all the \cuttlefish{} tuners as well as the local \texttt{State} of each worker. We use two layers of communication: lock-based synchronization of shared state within workers and asynchronous message rounds between workers. At every call to a tuner's \texttt{choose} method, the tuner first merges the local and non-local \texttt{State}s and then makes a decision using the merged state. At each call to \texttt{observe}, the tuner only updates the local \texttt{State}. Unsynchronized updates could override observations and prolong exploration, so on each worker we use lightweight locks to synchronize access to the \texttt{States} across threads, as indicated by the lock icon in~\Cref{fig:sharingfigure}. Workers send their local \texttt{State} to the centralized model store once every short interval (currently once every 500ms in our implementation). The centralized store saves the most recent local \texttt{State} that it has received from each worker. Then, the workers asynchronously pull from the model store a merged aggregation of the local \texttt{States} of all other workers. Upon receipt, this aggregated \texttt{State} becomes the worker's new non-local \texttt{State}. Workers may batch messages for multiple tuners at once.

This approach requires that tuning algorithms support associative, commutative merging of their learned state. \cuttlefish{} tuning algorithms do this by maintaining the sample statistics using one-pass, parallel algorithms~\cite{pebay2008formulas}.

This architecture has several desirable properties.
First, it is \textit{eventually consistent}: i.e.,
as long as the model store can communicate with all workers using an in-order channel, all observations are eventually available to all tuner instances. Second, if we assume physical operator runtime distributions do not vary across worker threads, the above eventually consistent guarantee is sufficient to reduce the theoretical behavior of this distributed tuning architecture to a centralized tuner with feedback delay.
Existing theoretical results on bandits with bounded delayed feedback~\cite{joulani2013online} suggest that our architecture will converge to the same tuning solution as a single centralized tuner with no feedback delay given enough tuning rounds. Third, if network partition occurs and a worker loses access to the central model store, the tuner will still eventually converge to the same solution as a centralized tuner.
Finally, we achieve these three properties using only a fixed memory overhead to maintain the states and without requiring blocking network calls. The overhead can be controlled by adjusting the frequency of asynchronous communication.

\section{Tuning in Dynamic Settings}
\label{sec:nonstationaryrewards}

So far we have assumed that the rewards distributions and workload properties do not vary across workers and over time, we now discuss how to tune operators when both change dynamically.
Such scenario arises, for instance, when 
the web crawl data from every website is stored as a single contiguous block on disk in the caption-generating model training workload mentioned in \Cref{sec:motivation}. The best operator variants to use may then change over time and differ across workers depending on the websites they happen to process. Such non-stationarity presents new challenges for tuning operators.

A number of techniques have been proposed for tuning for contextual and non-contextual MAB settings where rewards may change over time. These include using sliding windows of recent observations~\cite{garivier2011upper}, discounting older observations by a function of their age~\cite{garivier2011upper,burtini2015improving}, and resetting observations when a change in reward distributions is detected~\cite{hartland2006multi,allesiardo2015exp3}.

Another set of approaches, known as clustering of bandits, consider settings where multiple agents are solving different bandit problems over the same set of arms. The reward distributions of each arm may differ across problems. These techniques attempt to detect which agents are solving similar problems, i.e., problems with similar reward distributions. The agents solving similar problems then collaborate by sharing their  observations~\cite{gentile2014online,pmlr-v70-gentile17a,Korda:2016:DCL:3045390.3045528}. 

However, we are unaware of any technique for clustering of bandits when the agents may unknowingly change bandit problems over time. An agent currently solving a given tuning problem should use as many of the observations as possible that were generated from similar problems, regardless of which agent made the observations and when. This allows the agent to tune for the problem as quickly as possible. However, if an agent incorporates observations made for dissimilar tuning problems, the agent could incorrectly converge to poor tuning solutions. We design an approach to handle this setting and integrate it into \cuttlefish{}'s distributed shared-nothing tuning architecture.

One strawman approach to solve the clustering problem is as follows: we split the MAB rounds into multiple epochs, and each agent builds a separate observation state for every epoch. These states can be laid out in a two dimensional grid: one dimension is the epoch, and the other dimension specifies the agent that made the observations. We refer to these states as ``agent-epoch'' states. Each agent attempts to make decisions using all agent-epoch states that come from similar MAB problems as the agent's current agent-epoch state. In every MAB round, each agent uses a statistical similarity test to pairwise compare their current agent-epoch state with all other agent-epoch states. The agent aggregates their current agent-epoch state with all states that pass the test. The agent then uses this aggregated state to select an arm. Finally, when agents observe a reward, they will update only their current non-aggregated state, and they drop the aggregated state. At the start of every epoch, each agent begins a new agent-epoch state.

The above approach raises several challenges. (1) \textit{Growing overheads}: We need a principled method to discard or aggregate information from past epochs to avoid growing state overheads. (2) \textit{Statistical test for similarity}: If the test is too conservative, tuner instances will use only a fraction of all relevant observations. If the test is too liberal, tuner instances may use bad information that can interfere with their learning. In both cases, tuning quality will suffer. (3) \textit{Test \& aggregate overhead}: Identifying and aggregating similar states is expensive and requires communication. (4) \textit{Epoch length}: Short epochs enable faster reactions to changing conditions but add exploration overheads.

There are different approaches to limit memory overheads due
to state accumulation from past epochs. In \cuttlefish{}, we limit overheads by storing a single aggregated state per agent that combines all of the old epochs for that agent which are still relevant to its current state, and we store an additional state per agent for the current epoch. At the end of each epoch, the statistical test is used to check if the most recent epoch for every agent was similar to its aggregate of old epochs. If it is similar, the recent agent-epoch state gets merged into the aggregate of old epochs. If the statistical test fails, the aggregate state of old epochs is replaced with the recently-completed epoch's state.

To address the second challenge, \cuttlefish{} allows each tuner to specify an appropriate statistical test to check if two \texttt{State}s are similar. The non-contextual tuner we described in \Cref{sec:banditnocontext} uses a simplification of \textit{Welch's unequal variances t-test}~\cite{welch1947generalization} for this purpose. This test assumes two populations follow normal distributions with potentially unequal variances, and it checks if the two populations have equal means. The contextual tuner from \Cref{sec:contextualtuning} uses a statistical test from prior work on clustering contextual bandits~\cite{gentile2014online}. Additionally, agents may apply these statistical tests and aggregate similar states independently for every arm.
Finally, when observation states have too few observations for the statistical test to return confident results, the tests should always fail. Our evaluation in~\Cref{sec:dynamic-eval} shows that these statistical tests are empirically effective with this dynamic tuning strategy.

For the third challenge, we limit the overheads of state aggregation by only identifying and aggregating the similar states from non-local agents on the centralized model store, as part of the communication rounds in \cuttlefish{}'s distributed architecture. In our architecture, the central store receives two states per agent instead of just one state per worker: the aggregation of old epochs' states, and the state for the current epoch. The central store maintains both states, and sends pulling workers an aggregation of all non-local agent states which passed the statistical similarity test. The agent can then use this aggregation when choosing an arm at every round. \cuttlefish{} optionally allows the distributed architecture to treat every thread on every worker as a separate agent that maintains its own state objects rather than treating each multi-core worker process as a single agent and sharing state objects among threads within a single worker.

Finally, we allow \cuttlefish{} users to specify epoch length as a fixed number of tuning rounds, a fixed amount of wall clock time, or a single partition in the underlying data processing system. We recommend users match the epoch length to the frequency with which they expect workload changes to occur. We have implemented all of the above improvements and we evaluate them in the next section.

\section{Evaluation}
\label{sec:evaluation}

\begin{table*}[!t]
\centering
\begin{tabular}{llllll}
\\[-1.8ex]\hline
\hline \\[-1.8ex]
Operator    & Physical Operators & Dataset  & Workload Variants & Tuning Rounds \\ 
\hline \\[-1.8ex]
Convolution  & 3 Convolution Algs.   & 8091 Flickr Images (32 GB) & 3 Stationary, 3 Dynamic & 8091 (one per image) \\
Regex Matching  & 4 Regex Libraries   & 255,985 Common Crawl Docs (29.7 GB) & 8 Regexes & 255,985 (one per doc) \\
Parallel Join & 2 Local Join Algs.    & TPC-DS Data (about 200 GB) & 23 Queries & 512 (one per partition) \\
Simulated & 2 to 50 & --- & 14 Simulation Configs. & 50000 \\
\hline \\[-1.8ex]
\end{tabular}
\caption{Workload Summaries}
\label{tab:workloads}
\vspace{-0.3in}
\end{table*}

We have extended Apache Spark~\cite{zaharia2016apache} version 2.2 with a prototype of \cuttlefish{} such that Spark users can now directly insert tuners into the closures they pass to any of Spark's operators, including \texttt{map}, \texttt{flatmap}, and \texttt{reduce}. The resulting API is flexible enough to nest tuners inside physical variants of other tuners.

We evaluate \cuttlefish{}
on Amazon EC2 using r3.xlarge instances with 30.5 GB of memory and 4 virtual cores on an Intel Xeon E5-2670 v2 processor. We have workers communicate with the central model store once every 500 milliseconds.
We force class loading before all of our experiments to warm up the JVM. 

We use \cuttlefish{} to tune operators from our motivating example in \Cref{sec:motivation}: convolution, regular expression matching, and distributed parallel join. We call these \cuttlefish{}-tuned operators \textit{adaptive operators}. We also complement these operators with a synthetic one that enable us to vary additional parameters. 
\Cref{tab:workloads} summarizes the operators and the input
data that we use in the experiments.

We implement the three convolution variants described in \Cref{sec:usingcuttlefish} and test the adaptive operator on 8091 3-channel images from the Flickr image hosting website~\cite{hodosh2013framing} with different randomly generated filters. In the stationary workloads, there is an equal probability of convolving an image with any of the filters in the given set. In the dynamic workloads, we vary the probability of picking a given filter set for each image over time and across the threads in the cluster.

We implement four versions of regular expression matching
using four different libraries.\footnote{Java TCL library (https://github.com/basis-technology-corp/tcl-regex-java), Jakarta ORO library (https://attic.apache.org/projects/jakarta-oro.html), JRegex (http://jregex.sourceforge.net), and the regex utilities built into the standard Java distribution.} To test the adaptive operator, we use eight different highly-rated regex queries from the regex sharing tool RegExr\footnote{http://regexr.com} to search through a contiguously-stored sample of approximately 256 thousand internet web pages collected by the Common Crawl project.\footnote{http://commoncrawl.org}

For the join, we extend Spark SQL~\cite{armbrust2015spark} with the adaptive distributed join operator from \Cref{sec:tuningpipelined}, which first hash-partitions input relations across a cluster and then, per-partition, picks either a local hash or sort-merge join. We test this operator on the TPC-DS benchmark~\cite{nambiar2006making} with a scale factor of 200. We use this operator in all equality joins over relations that Spark SQL's query optimizer decides are too large to broadcast, as the optimizer lacks heuristics to choose a non-broadcasting join and just picks whichever join is hard-coded in the Spark SQL configuration parameters. \cuttlefish{} tunes each join independently. We allow Spark SQL's optimizer to select the build relation for local hash joins using the same rules it uses for shuffled hash joins. We configure Spark SQL queries to use 512 data partitions in all joins and shuffles, providing 512 tuning rounds for each join.

\subsection{Overall Operator Tuning Performance}
\label{sec:overall-op-evaluation}

In this section, we report the throughput of our three adaptive operators on all of the stationary workload variants, with a cluster of 8 AWS r3.xlarge instances and 32 total virtual cores.

\begin{figure}[!t]
  \centering
  \includegraphics[width=1.0\columnwidth]{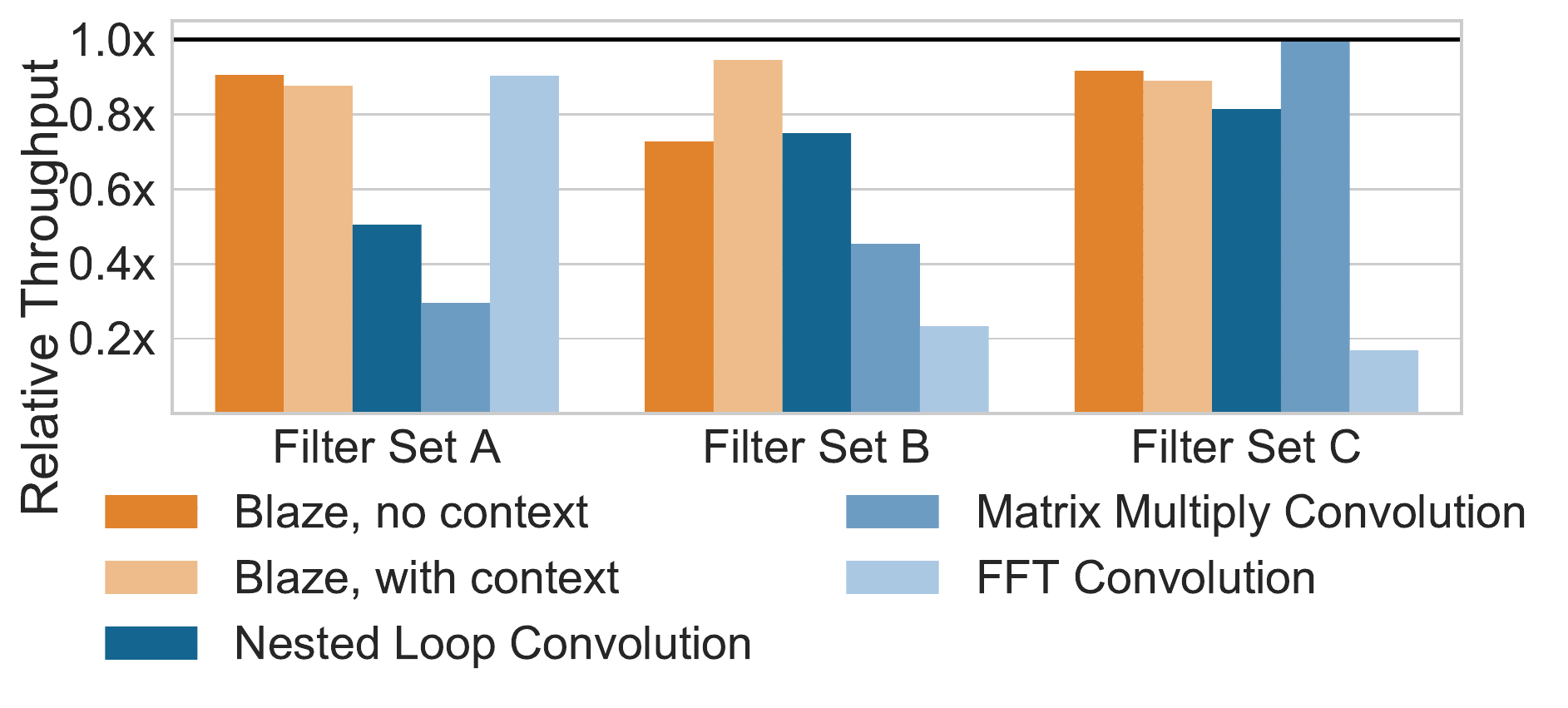}
  \vspace{-0.3in}
\caption{Convolution relative throughput}
\label{fig:flickr_8node}
\vspace{-0.2in}
\end{figure}

\noindent\textbf{Convolution.}
We use three sets of stationary filters. Set A comprises five $25\times25\times3$ pixel filters per image. Set B applies to each image a random set of 1 to 25 equal-dimension filters between $5\times5\times3$ pixels and $30\times30\times3$ pixels. Set C uses 50 $8\times8\times3$ pixel filters. With the 8-node cluster size, there are approximately 253 images per core, a total of 253 tuning rounds per thread.

We show the relative throughput of the contextual and context-free adaptive operators as compared to the three baseline convolution algorithms in \Cref{fig:flickr_8node}. The contextual convolution operator uses four context features: the total number of pixels in the image, the total number of pixels in the set of filters, and two additional terms that are in the asymptotic computational complexity of the FFT. We normalize ``$1.0\times$'' throughput to a hypothetical ``ideal'' oracle that perfectly picks the fastest physical operator for every convolution, where it can choose a different algorithm for different images and filters.

\Cref{fig:flickr_8node} shows that the best single-choice baseline
varies between workloads and rarely matches the maximum throughput of the
oracle. Importantly, as seen with filter set C, the fastest single convolution algorithm (in this case matrix multiply) can be as much as $6.25\times$ faster than the slowest convolution algorithm. This is important as 
\Cref{fig:flickr_8node} also shows that the context-free adaptive operator achieves 91\% to 99\% of the throughput of the fastest single operator variant, even with the necessary exploration rounds and the dramatic performance differences between algorithms observed in Set C. Finally, we observe that contextual tuners require more exploration and perform worse than the context-free operator when there is not much to be gained by contextually picking operator variants (Set A and C). With Set B, the constantly changing convolution kernels means that learning when to select each algorithm with context significantly outperforms context-free tuning, as well as all of the individual convolution algorithms. 

\begin{figure}[!t]
  \centering
  \includegraphics[width=1.0\columnwidth]{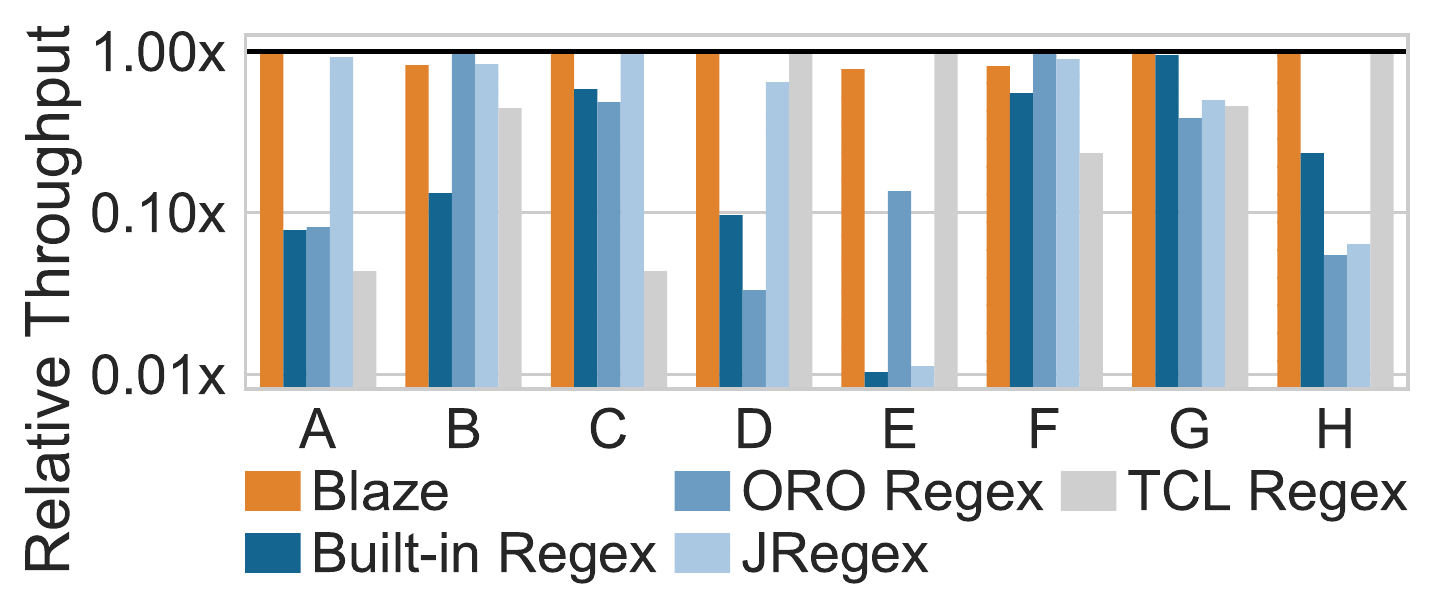}
  \vspace{-0.2in}
\caption{Regular expression throughput on Common Crawl data. Y-axis uses
a log scale. Regex A matches any URL, B matches all three-word trigrams, C matches HTML hyperlinks, D matches phone numbers, E matches valid emails, F matches prices specified in US currency, G matches all CSS color definitions, and H matches valid IPv4 addresses}
\label{fig:regex_8node}
\vspace{-0.2in}
\end{figure}

\noindent\textbf{Regex Matching.}
\Cref{fig:regex_8node} shows the relative throughput results for the regex matching operator. Because modeling regular expression performance requires different complex features for each regex (e.g., how many times different character subsequences in the regex appear in each document), we do not use a contextual tuner. We normalize the relative throughput against the single fastest regex library for each query. As~\Cref{fig:regex_8node} shows, the
regex matching throughput varies dramatically between libraries and different libraries perform best on different queries.
For example, the fastest library when processing Regex E has almost $105\times$ higher throughput than the slowest library. Additionally, the runtime distributions for each library have a massive spread, even for a single regular expression. With Regex 5, the built-in java Regex utilities takes only 33$\mu$s to process the fastest document, but the slowest document search takes over 1000s, a difference of over 8 orders of magnitude. 
These large runtime differences across documents in a single query mean that \cuttlefish{} requires significant exploration to confidently identify which library to use. Yet, accidentally making the wrong decision during exploration can have long-lasting consequences for the cumulative throughput. Despite these challenges, 256 thousand documents across our 8 node cluster provide sufficient tuning time and enable \cuttlefish{} to reach over 99\% of the throughput of the fastest single regex library for many of the queries and over 72\% for all.

\begin{figure*}[!t]
  \centering
  \includegraphics[width=1.0\textwidth]{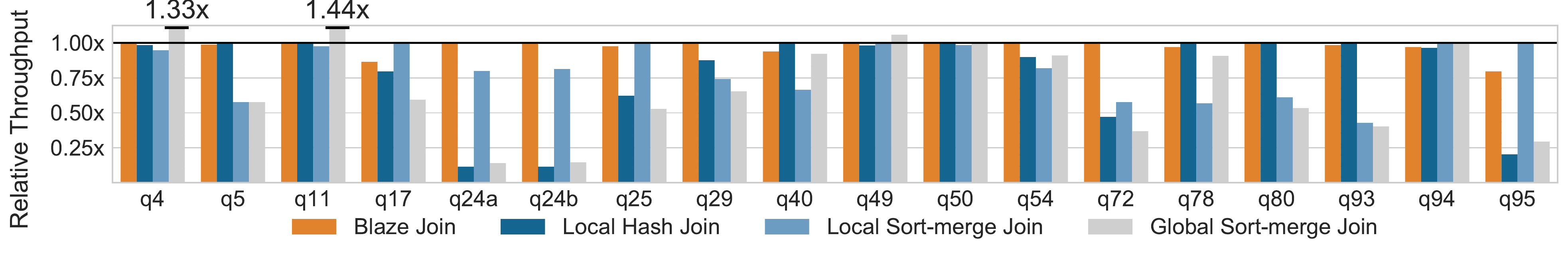}
  \vspace{-0.3in}
\caption{TPC-DS Join throughputs}
\label{fig:tpcds_8node}
\vspace{-0.2in}
\end{figure*}

\noindent\textbf{Parallel Joins.}
We evaluate the context-free adaptive join operator, with 512 partitions per join spread across the cluster. We choose this number because using too many partitions or too few can harm Spark SQL's performance. We do not test a contextual variant because Spark SQL does not expose per-partition selectivity and cardinality estimates, and computing these statistics would require expensive passes over intermediate query results.
\Cref{fig:tpcds_8node} shows the results for the TPC-DS queries that can utilize the adaptive join in one or more logical joins (e.g., q72 contains 3 equality joins which we can replace with our adaptive variant). We
measure and compare only the time spent in the join operators.
We compare the relative join throughput with \cuttlefish{} to the throughput when always using one of the two partitioned local joins, normalizing against the fastest of the two for each query. We also compare the join throughput to that of a global sort-merge join, Spark SQL's default choice for all of these joins.

The greatest challenge in this experiment is that there are only 16 partitions per core, which leaves little time for the \cuttlefish{} tuner to explore. However, \Cref{fig:tpcds_8node} shows that for all the TPC-DS queries, the adaptive operator is still able to stay within 80\% of the throughput of the better of the two local join variants. In q24a, q24b, q29, q54, and q72, \cuttlefish{} actually outperforms both picking a local hash join and picking a local sort-merge join for every partition. This is because these queries have multiple logical joins that we replace with the adaptive join operator, and \cuttlefish{} identifies a different join strategy for each logical join. This outperforms selecting the same implementation for every non-broadcasted equality join.

In all queries except for q4, q11, and q49, the join throughput of our adaptive operator matches or outperforms the Spark SQL's optimizer's choice of a global sort-merge join by up to 7.5$\times$ (for q24). Even for those three queries, our adaptive join achieves over 70\% of the join throughput of the global sort-merge join.

\subsection{Simulation: When does tuning work best?}

\begin{figure}[!t]
  \centering
  \includegraphics[width=1.0\columnwidth]{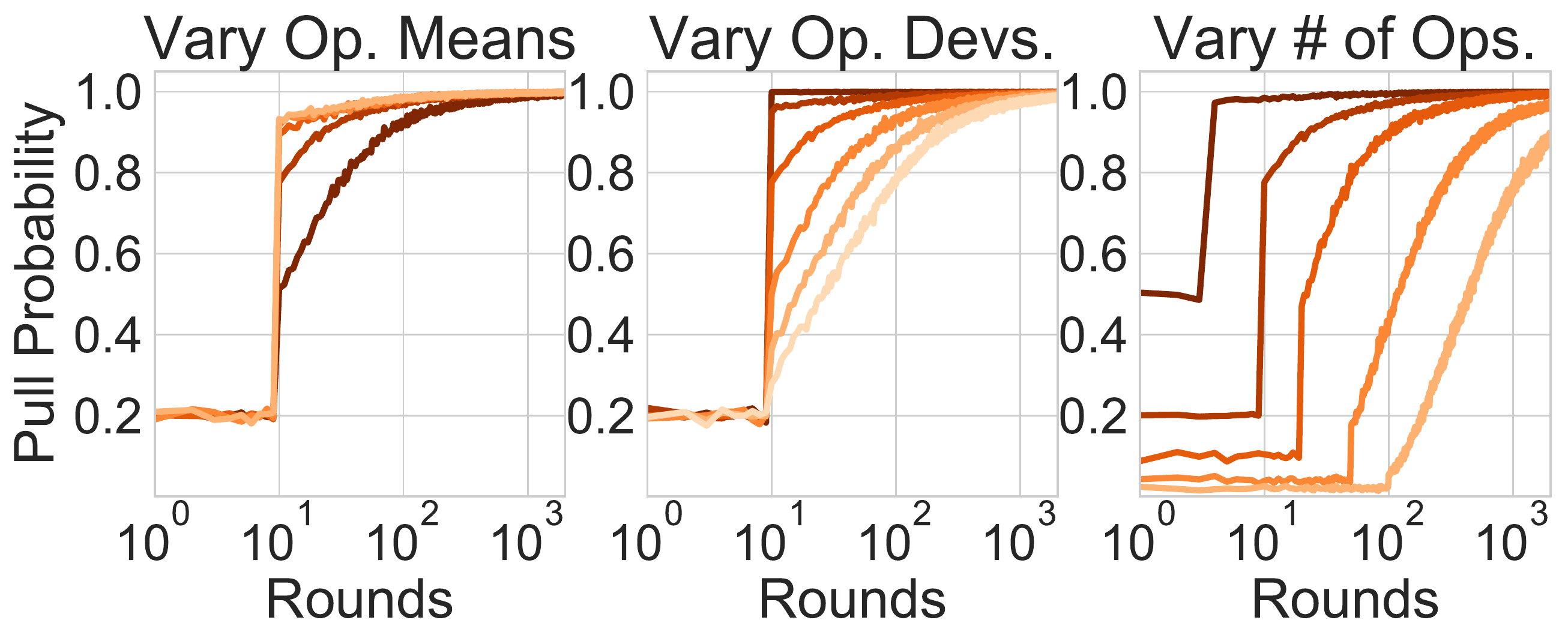}
   \includegraphics[width=1.0\columnwidth]{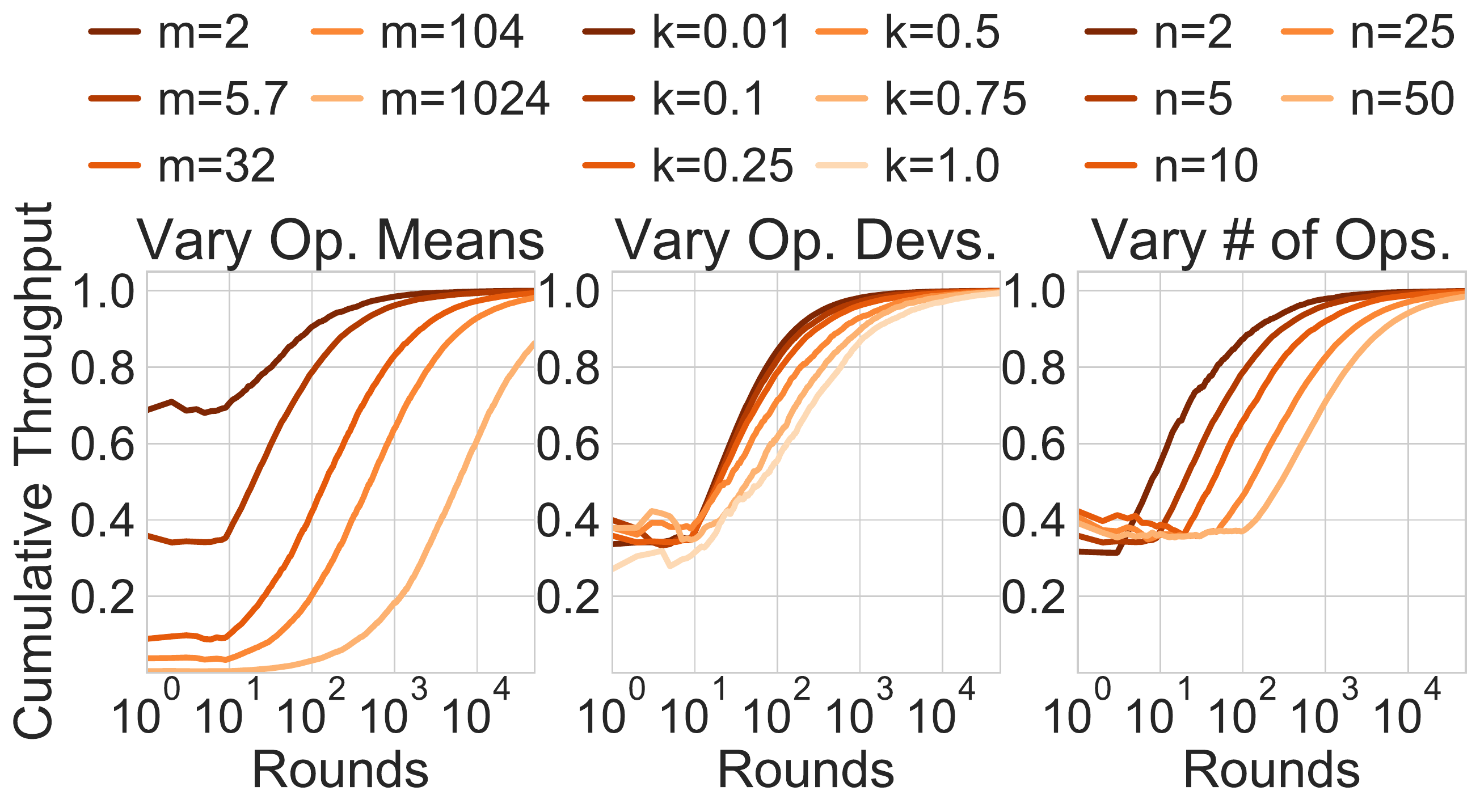}
  \vspace{-0.3in}
\caption{Top: Probability of picking the fastest variant. Bottom: Cumulative throughput. Both share a legend.}
\label{fig:pull_probability}
\vspace{-0.2in}
\end{figure}

To study \cuttlefish{}'s performance in more detail, we use \cuttlefish{} to tune a synthetic no-op operator with variants that have pre-defined runtime distributions. We configure a single-threaded simulation as follows. We create $n$ operator variants with Gaussian runtime distributions. We set the mean runtime $\mu$ of the fastest variant to $1$ time unit and set $\mu$ for the slowest variant to $m$ time units. We space $\mu$ for the other variants exponentially within this range. Finally, we set the standard deviation of each variant's distribution to $\sigma=\mu\cdot k$, where $k$ is a scalar constant. We set the default simulation configuration to tune between $n=5$ operator variants, set the slowest variant to $m=5.7\times$ slower than the fastest variant, and set the standard deviation multiplier $k=0.25$. We individually vary each of these parameters while holding the others at the defaults. We test $n=2$ to $n=50$ operator variants, $m=2$ to $m=1024$ time units, $k=0$ to $k=1$ as the standard deviation multiplier. We then allow \cuttlefish{} to tune these operators using its default context-free tuner for 20,000 rounds using 200 trials for each configuration. We record both the instantaneous probability that \cuttlefish{} will choose the fastest variant at each tuning round and the cumulative throughput after a given number of rounds. We show the results of these simulations in \Cref{fig:pull_probability}. The x-axes of these plots are all in log-scale.

As \Cref{fig:pull_probability}(top) shows, for all configurations, the probability of picking the fastest variant
is low at first as the context-free MAB tuning algorithm tries each arm
two times. When we vary the magnitude of $m$, the speed gap between variants, we observe that close operator runtimes require \cuttlefish{} to perform more exploration before it confidently identifies the fastest variant.
Similarly, tight runtime distributions (low runtime standard deviations) enable \cuttlefish{} to converge faster to the best arm than
distributions with wide runtime spread.
Finally, the top-right plot shows that increasing the number of arms increases the minimum required amount of exploration. Although the curves in this plot appear almost parallel to each other, because the x-axis is in log scale this plot also suggests that increasing the number of arms causes \cuttlefish{} to explore significantly more even after it completes the minimum amount of exploration.

The three plots in \Cref{fig:pull_probability}(bottom) show the relative cumulative throughput after every round as we vary the simulation parameters. The most important thing to notice is that the x-axis now goes from 1 to 20,000 in log scale, while in \Cref{fig:pull_probability} the scale only went from 1 to 2000. Even when \cuttlefish{} converges to the fastest operator variant in a very small number of tuning rounds, the cumulative throughput takes orders of magnitude longer to recover from the slow variants tried during exploration. This is best highlighted by comparing the top-left and bottom-left plots, which show the probability of picking the fastest variant as we increase the gap between the slowest and fastest operator variants, and the resulting cumulative throughput. Although making the slower variants orders of magnitude slower allows \cuttlefish{} to converge to the fastest variant in even fewer rounds, it dramatically harms the cumulative throughput. For example, when m=2 time units the synthetic operator takes 100 rounds to reach a cumulative throughput of 0.9 operations per time unit, but when m=1024 the tuner reaches takes 10,000 tuning rounds to reach a cumulative throughput of just 0.6 operations per time unit. The bottom-middle plot shows that increasing the standard deviation of each operator variants' runtime distribution also slows down the improvement in cumulative throughput due to requiring more exploration, but the effect is far less dramatic. However, changing the number of variants also has a dramatic impact on the number of rounds the cumulative throughput requires to improve. For example, when there are only n=2 variants the synthetic operator reaches a cumulative throughput of 0.6 after just over 10 rounds, but when n=25 variants the synthetic operator takes over 100 rounds to reach that same cumulative throughput.

We show the performance of our real operators after each round overlaid on top of the above synthetic
experiments in the Appendix.

\subsection{Tuning with Context}
\label{sec:evalcontext}

\cuttlefish{} supports contextual tuning, but it relies on users to provide the context features. Identifying good
features can be challenging to users. In this section,
we evaluate \cuttlefish{}'s sensitivity to feature quality. We use the convolution operator because it has
an easy to identify and extract set of four ``good'' features. The first two are the number of pixels in an image ($n$) 
and the number of pixels in the filters ($k\cdot m$, where $k$ is the number of filters to convolve the image with and $m$ the number of pixels in each filter). Additionally, from the $O(N\log N)$ computational complexity of the FFT, we get the features $n\log n$ and $k\cdot m\log m$. We further add four ``random'' features to test the tuning algorithm, which we generate for each convolution from standard Gaussian distributions. The contextual operator does not require any explicit bias feature due to the built-in feature pre-processing (described in Appendix A).

\begin{figure}[!t]
  \centering
  \includegraphics[width=1.0\columnwidth]{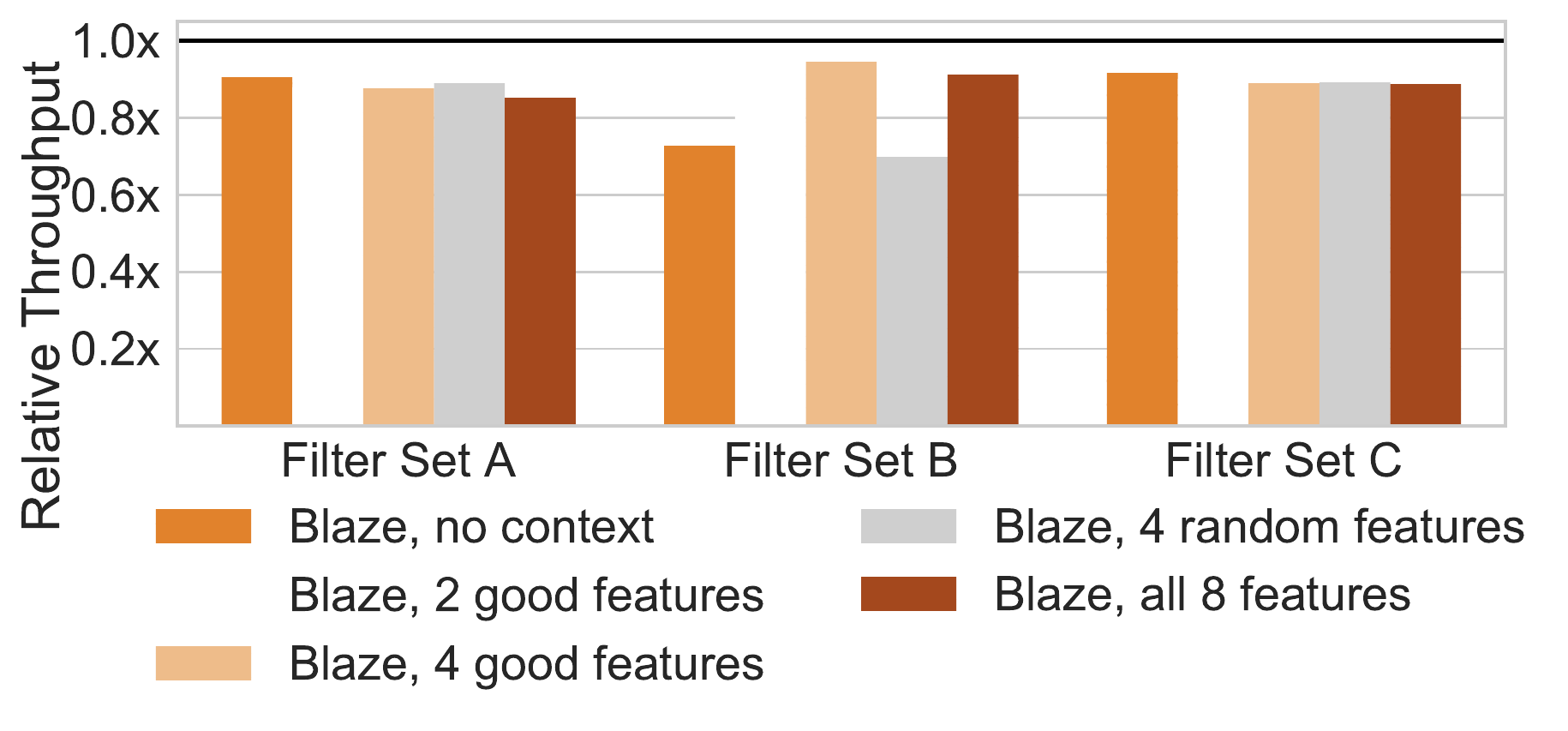}
  \vspace{-0.3in}
\caption{Contextual convolution with various features}
\label{fig:contextcomparison}
\vspace{-0.2in}
\end{figure}

\Cref{fig:contextcomparison} shows the results. 
We show the relative throughput using the same normalization as \Cref{fig:flickr_8node}.
As the figure shows, when we convolve the Flickr images using filter sets A and C, the contextual cost model is insufficient to provide any benefit over the context-free adaptive operator, and every added feature increases the necessary amount of exploration.
With filter set B, \cuttlefish{} significantly benefits from the good features and is resilient to the addition of bad features, unless they are
the only features given to \cuttlefish{}. Even in that case, \cuttlefish{}'s performance remains close to its performance without any context at all.

To measure the computational overhead of adding extra features, we execute synthetic microbenchmarks with unoptimized \cuttlefish{} tuner implementations. We vary the number of features and measure the time it takes tuners to make decisions. We find that context-free tuners with five operator variants take $0.03$ms per choose-and-observe round. Contextual tuners with five operator variants and 2, 4, and 8 features respectively take $0.034$, $0.046$, and $0.082$ms per round. The overhead is thus visible but remains small in absolute value even with 8 features. We further explore \cuttlefish{}'s system overheads in Appendix D.

\subsection{Scaling and the Distributed Architecture}

\begin{figure}[!t]
  \centering
  \includegraphics[width=1.0\columnwidth]{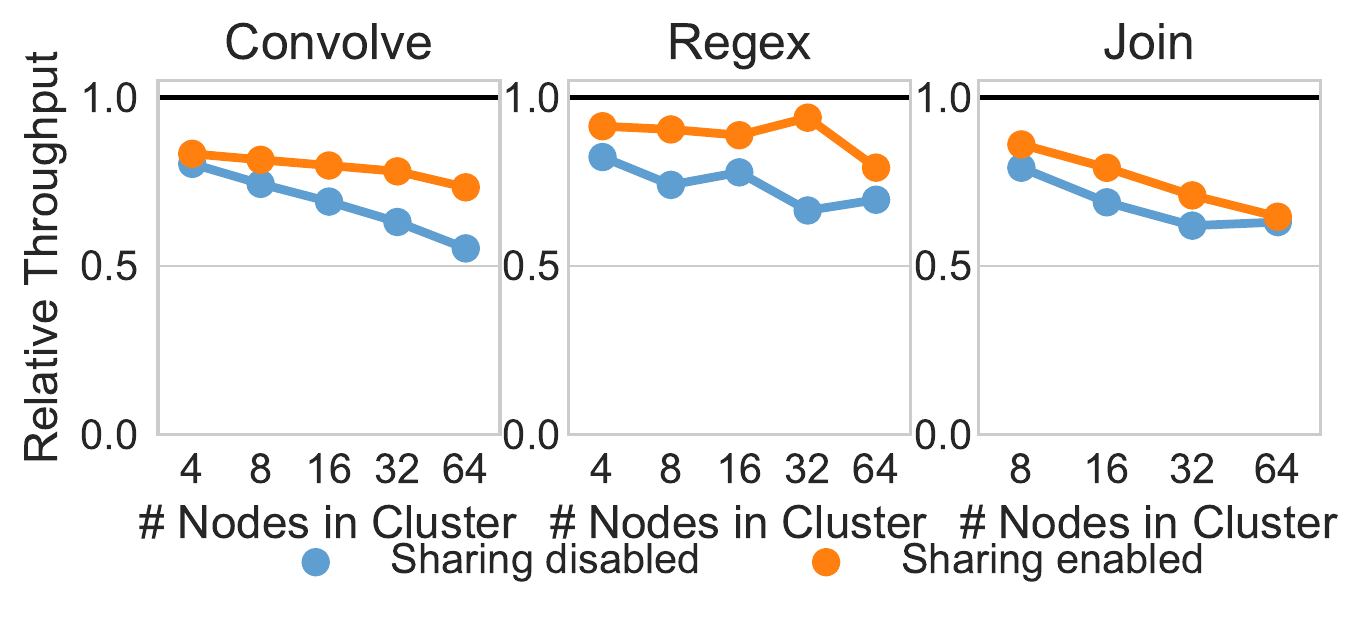}
\caption{\cuttlefish{} at different cluster sizes}
\label{fig:scaling_all}
\end{figure}

Next, we measure \cuttlefish{}'s tuning performance as we scale to larger cluster sizes. We test the context-free adaptive operators on our three real datasets. We test \cuttlefish{}'s distributed tuning architecture, and compare it to an architecture that tunes each thread independently. We set \cuttlefish{}'s distributed architecture to communicate between worker nodes and the master once every 500ms. \Cref{fig:scaling_all} shows the geometric means of the relative throughputs for all the workload variants for each dataset. We test the convolution and the regex operators with cluster sizes from 4 to 64 instances. Unfortunately, Spark SQL crashes when executing the TPC-DS benchmark at our scale factor on the four node cluster, so we only test the join operator on 8 to 64 node clusters. As the figure shows, for convolution and regex, \cuttlefish{}'s approach, which includes information sharing across independent workers, successfully
limits the impact of cluster scaling on tuning. In the absence of sharing, all workers explore in parallel and thus more work is wasted to exploration rather than exploitation. For joins, cluster size impacts performance even with sharing enabled. Because it has only 512 tuning rounds across the entire cluster, that leaves only two tuning rounds per core on the 64 node cluster, so the parallel join finishes before the tuners can exploit what they have learned from exploration, or to share any observations.

\subsection{Dynamically-Changing Environments}
\label{sec:dynamic-eval}

\begin{figure}[!t]
  \centering
  \includegraphics[width=1.0\columnwidth]{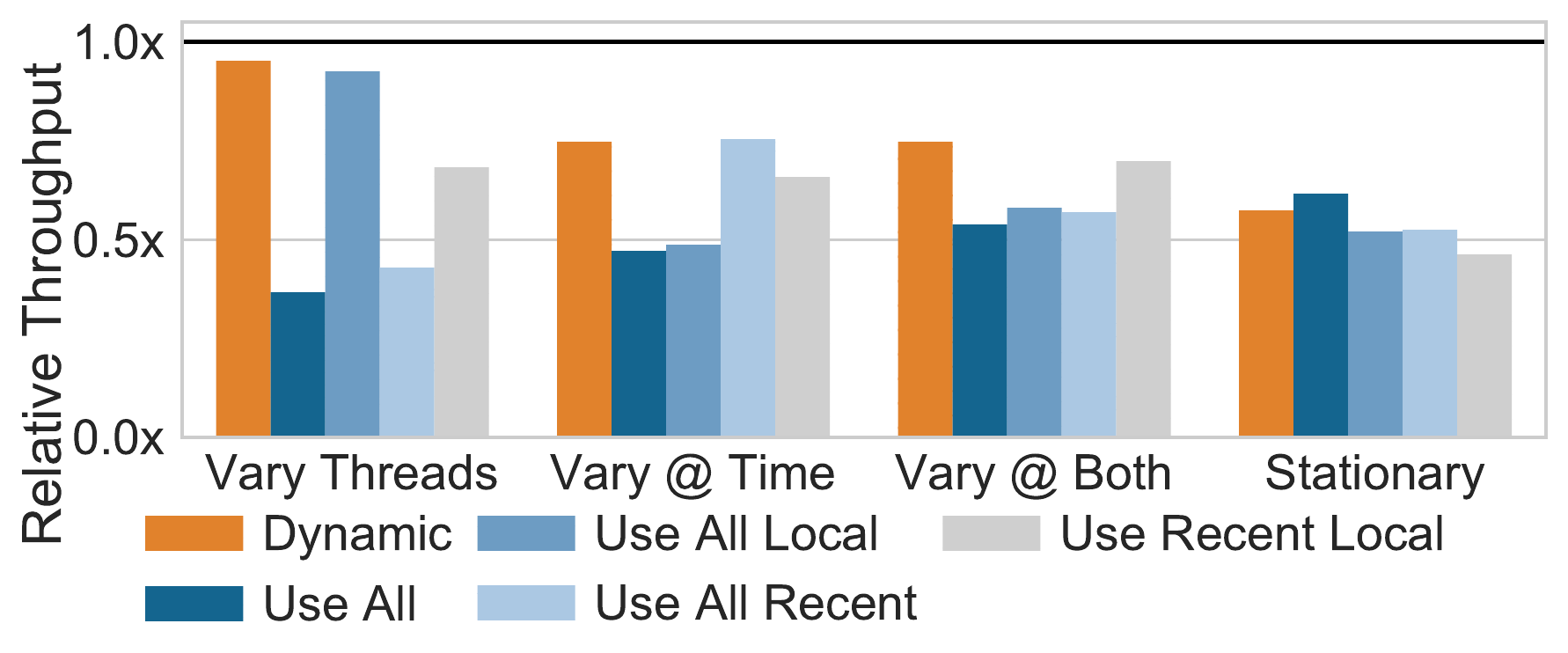}
\caption{Convolution in Dynamically-Changing Settings}
\label{fig:flickr_8node_nonstationary}
\end{figure}

To explore how well \cuttlefish{} works in dynamic environments, we test the adaptive convolution operators on the Flickr dataset using workloads that change convolution filters across threads and over time. We test \cuttlefish{}'s dynamic tuning approach from \Cref{sec:nonstationaryrewards} using one agent per core in the cluster, with the epoch length set to 15s. We compare it to four control strategies: \cuttlefish{}'s default non-dynamic distributed architecture from \Cref{sec:sharednothing} that uses all observations; tuning threads using all local observations without sharing across threads; tuning threads using only observations from the most recent epoch but sharing across threads; and tuning threads using only observations from the most recent epoch without sharing across threads.

\Cref{fig:flickr_8node_nonstationary} shows the relative throughputs for the context-free adaptive convolution when using 4 dynamic workload variants:
``Vary Threads'' assigns a different group of filters per thread (with different counts and dimensions for each group). But, it keeps the group of filters constant for all images processed by that thread. ``Vary @ Time'' randomly switches the group of filters at different points in time, but at any given time it uses the same group of filters for all images being convolved across the whole cluster. ``Vary @ Both'' uses different groups of filters for different threads, and it randomly switches the group of filters over time for all threads. Finally, `Stationary' is a workload with similar properties to filter set B from the non-dynamic convolution experiments. For each image, it selects a different-sized group of filters, but the probability of selecting each group does not change over time or across threads in the cluster.

Each control strategy performs best in a different workload variant. The dynamic tuner matches or outperforms the controls in each of the three dynamically varying environments. In these settings, it significantly outperforms \cuttlefish{}'s default non-dynamic tuner that uses all observations. However, the dynamic \cuttlefish{} tuner performs slightly worse than \cuttlefish{}'s non-dynamic tuner in the stationary setting, where operator runtime distributions remain constant. This is because the dynamic one-agent-per-core tuners have two main overheads over the default non-dynamic distributed tuners: First, \cuttlefish{}'s non-dynamic distributed architecture immediately shares observations across cores without waiting for communication rounds. Next, at the start of each epoch dynamic tuners experience a short mandatory exploration period before they can utilize observations from other agents and epochs.

In Appendix C we observe that contextual tuners can also benefit from the dynamic tuning approach described in \Cref{sec:nonstationaryrewards}, except when the context can model most or all of the dynamic changes in the workloads.


\vspace{-0.1in}
\section{Related Work}

\textbf{Adaptive query processing:} \cuttlefish{} is one of many adaptive query processing techniques that modify query execution plans at runtime~\cite{deshpande2007adaptive, babu2005adaptive, hellerstein2000adaptive}. In contrast to query re-optimization~\cite{ng1999dynamic}, which re-generates entire query plans using improved statistics when optimization mistakes are detected at runtime, \cuttlefish{} continuously updates physical operator implementations. Other adaptive approaches do continuously modify query plans~\cite{rundensteiner2004cape}, but unlike \cuttlefish{}, these still rely on explicit query planner heuristics and cost models.
Eddies~\cite{avnur2000eddies} and related tuple-routing approaches~\cite{babu2004adaptive, li2007adaptively} adaptively learn relational operator orderings. In contrast, \cuttlefish{} changes operator implementations and supports
non-relational operators. Most closely related to \cuttlefish{}, Micro Adaptivity in Vectorwise~\cite{ruaducanu2013micro} uses epoch-based $\epsilon$-greedy bandit policies to pick between several black-box ``flavors'' of vectorized operators. However, the Vectorwise approach requires hyperparameters to be set, it does not explore contextual tuning, it is limited to operators with a vectorized execution model, and although it is built in a parallel DBMS the paper only explores a single-core setting.

\textbf{Adaptive operators: } Several adaptive operators have been developed in the literature including adaptive scans~\cite{borovica2015smooth}, aggregates~\cite{muller2015cache}, and others~\cite{rundensteiner2004cape}. These approaches, however, still rely on explicit rules and cost models that developers must provide.

\textbf{Autonomic databases:} Self-tuning database systems explore how to automatically materialize views, build indices, and modify DBMS configuration parameters~\cite{chaudhuri2007self}. Recently, 
\textit{Ottertune}~\cite{van2017automatic} showed that Gaussian Process bandits can effectively tune the high-dimensional space of DBMS configuration parameters to optimize various query execution metrics. 
In contrast, our work focuses on tuning physical operator implementations.


\textbf{Offline \& Online Autotuners:}
\textit{Petabricks}~\cite{ansel:pldi:2009} and other offline tuners provide libraries that programs can use to easily support offline tuning during compilation time. \textit{Opentuner}~\cite{ansel2014opentuner} in particular uses bandit techniques to select program mutations during tuning. 
Offline tuners, however, require representative tuning workloads and a dedicated period of time to profile and tune the programs offline, which
our approach does not require.
SiblingRivalry~\cite{ansel2012siblingrivalry} explores tuning programs online using dueling bandit competition-based policies on multi-core machines. However, this competition based strategy requires dedicating half of all processor resources to speculative execution. Also, SiblingRivalry cannot not share learning across machines, and it does not support contextual tuning.

Bandits have also been used to tune software applications online. REX 
\cite{199370} showed that linear Thompson sampling can be used to dynamically swap component modules online in software applications and successfully tuned a web server. Pytheas~\cite{jiang2017pytheas} demonstrated that bandits can be used to tune networked applications, such as VOIP and video streaming, according to features of the network. Additionally, REX does not support distributed tuning, while applications can only tune using Pytheas via calling external network calls to Pytheas servers. In contrast, \cuttlefish{} focuses on tuning operators in data processing workloads, tuning at much finer sub-millisecond timescales at the application servers, and still managing to share bandit learning across machines.

\textbf{Reinforcement learning} has been successfully used in various tuning scenarios for database systems in particular resource allocation and query scheduling~\cite{DBLP:conf/sigmod/OrtizLB16,DBLP:conf/sigmod/KonstantinouATBKS12,DBLP:conf/ccgrid/TsoumakosKBSK13,DBLP:conf/cidr/MarcusP17}. In contrast, \cuttlefish{} is a new approach to physical operator tuning.

\section{Conclusion}

\cuttlefish{} is a lightweight online tuning primitive that enables developers to build adaptive operators at any granularity in a physical query plan. \cuttlefish{} tuners choose between physical operator variants during query execution. They use Thompson sampling multi-armed bandit heuristics to balance exploration and exploitation of the operator variants and perform efficient tuning even in shared-nothing distributed settings. They can further learn contextual cost models and can react to changing conditions. Our experiments 
show join throughput improvements of up to 7.5$\times$ compared with Spark SQL's query optimizer. \cuttlefish{}'s adaptive operators can also reach 72-99\% of the throughput of an oracle that always selects the optimal algorithm, even when suboptimal implementations are significantly 
slower than the optimal.

\begin{acks}
We would like to thank Kevin Jamieson and Sham Kakade for helpful discussions about the idiosyncrasies of multi-armed bandit theory and its applicability to this setting. We would also like to thank Calvin Loncaric, Niel Lebeck, Parmita Mehta, Laurel Orr, Jennifer Ortiz, Guna Prasaad, and Dan Suciu for providing feedback on earlier drafts. This work is supported in part by the National Science Foundation under grants IIS-1247469, IIS-1546083, IIS-1651489, OAC-1739419, \& CNS-1563788; DARPA award FA8750-16-2-0032; DOE award DE-SC0016260; the Intel-NSF CAPA center; the Intel Science \& Technology Center for Big Data, a Weil Family Endowed Fellowship in Computer Science \& Engineering, and gifts from Adobe, Amazon, and Google. Tomer Kaftan is supported by the NSF Graduate Research Fellowship.

\end{acks}

\bibliographystyle{ACM-Reference-Format} 

\bibliography{sigproc} 

\section*{Appendix}

\subsection*{A. Details of Contextual Learning}

\begin{figure}[t]
\centering
\begin{lstlisting}[language=python, mathescape]
# To sample from arm A's posterior expected reward distribution, given context vector x:

# Let X represent the observed context vectors for arm A
# Let y represent the observed reward scalars for arm A
# Let $\lambda$ be a regularization parameter > 0
# Let $I$ be the identity matrix of size dim(x) by dim(x)

model_mean = (corr(X, X) + $\lambda I$ / len(X))$^{-1}$corr(X, y)
model_covariance = (corr(X, X) + $\lambda I$ / len(X))$^{-1}$
sampled_model = multivariate_gaussian(
    mean = model_mean,
    covariance_matrix = model_covariance / len(X)
).sample()

standardized_x = (x - mean(X)) / diag(cov(X, X))
r = standardized_x^T*sampled_model
sampled_reward = r * sqrt(var(y)) + mean(y)

return sampled_reward
\end{lstlisting}
\caption{Contextual Tuning algorithm}
\label{fig:contextthompsonsampling}
\end{figure}

\cuttlefish{}'s default contextual tuner when users provide context features combines \textit{Thompson sampling with linear payoffs} (linear Thompson sampling)~\cite{agrawal2013thompson} with a pre-processing step that \textit{standardizes} the rewards and context vectors. Linear Thompson sampling learns a regularized linear model for the rewards of each arm given a context vector.
However, because we do not expect \cuttlefish users to spend extra execution passes on feature engineering and pre-processing, many of their features will be of different magnitudes, or correlated with other features. Linear model training algorithms work less effectively with such features~\cite{marquardt1975ridge}. To train cost models more resiliently, we thus integrate automatic feature pre-processing into our single online tuning pass. Our pre-processing step \textit{standardizes} the rewards and context vectors by centering and scaling the context vectors such that for every operator variant the sample mean of the context vectors is the zero vector, and the sample variance is a vector of all ones. We also center and scale the rewards of each operator variant. To combine the standardization step with the model learning in a single online pass, we use numerically stable, single-pass computations for the mean, covariance, and correlation~\cite{pebay2008formulas} of the contextual features and the rewards. We show the pseudocode of the extended approach with standardized contextual features in the Appendix.

\Cref{fig:contextthompsonsampling} shows how Linear Thompson sampling samples from the posterior distributions over the expected rewards of every operator variant. Linear Thompson sampling does this sampling when it \texttt{chooses}, and as in the previous subsections this Thompson sampling tuner picks the operator variant with the highest sampled expected reward. Line 8 learns a best-fit regularized linear cost model given the observed standardized contexts and rewards, and Line 9 computes the covariance of the model parameters. In lines 10 to 13, the tuner samples a candidate cost model from the posterior distribution over the possible linear cost models. This distribution takes the form of a multivariate Gaussian centered at the best-fit model, and whose covariance is the model parameter covariance divided by the number of observations. The tuner then standardizes the current context vector in Line 16, and in Line 17 it predicts an expected standardized reward for the current context vector using the candidate cost model. Finally, in Line 18 the tuner un-standardizes the predicted expected reward, so that it can compare the sampled expected rewards across all the operator variants.

\subsection*{B. Real vs Synthetic Operator Tuning}

\begin{figure}[!t]
  \centering
  \includegraphics[width=1.0\columnwidth]{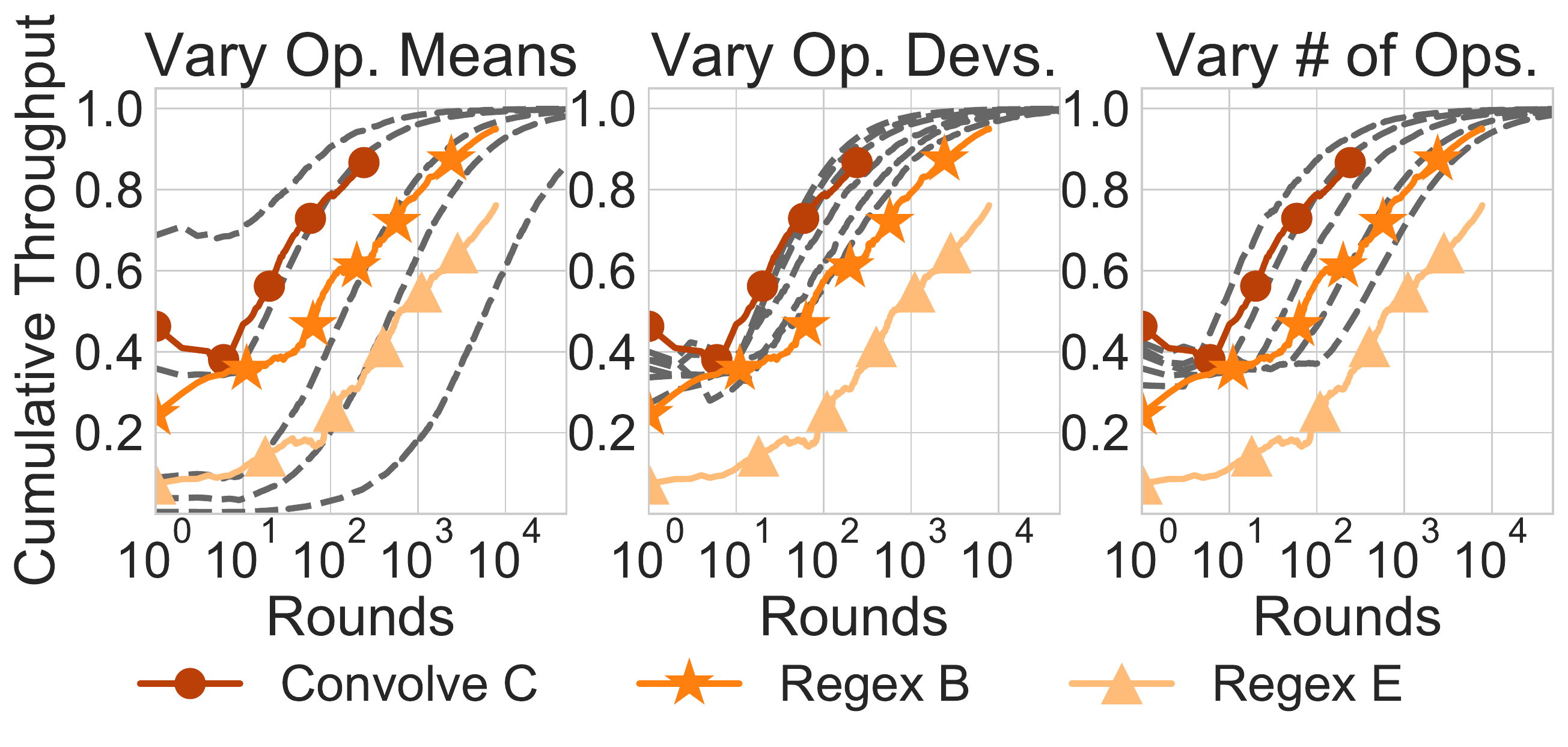}
\caption{Real operator throughput after each tuning round overlaid on synthetic results from \Cref{fig:pull_probability} (bottom)}
\label{fig:cumulative_reward_overlay}
\end{figure}

We test how closely our synthetic simulations from the evaluation match the results of our real adaptive operators. We test the adaptive operators from some of our earlier workloads after disabling the distributed architecture and making the \cuttlefish{} tuners tune each thread independently. In \Cref{fig:cumulative_reward_overlay}, we overlay these results on the cumulative throughput simulation from \Cref{fig:pull_probability} (bottom). We show the adaptive convolution operator with filter set A, and the adaptive regex operator with regexes B and E. We do not overlay any adaptive TPC-DS join query, because they have only 16 tuning rounds per thread, which is too few to meaningfully compare to the simulated results. The cumulative throuputs of all three workloads are in line with what the simulated results suggest, given the runtime properties of the operator variants in each workload.

\subsection*{C. Contextual Operators in Dynamic Workloads}

\begin{figure}[!t]
  \centering
  \includegraphics[width=1.0\columnwidth]{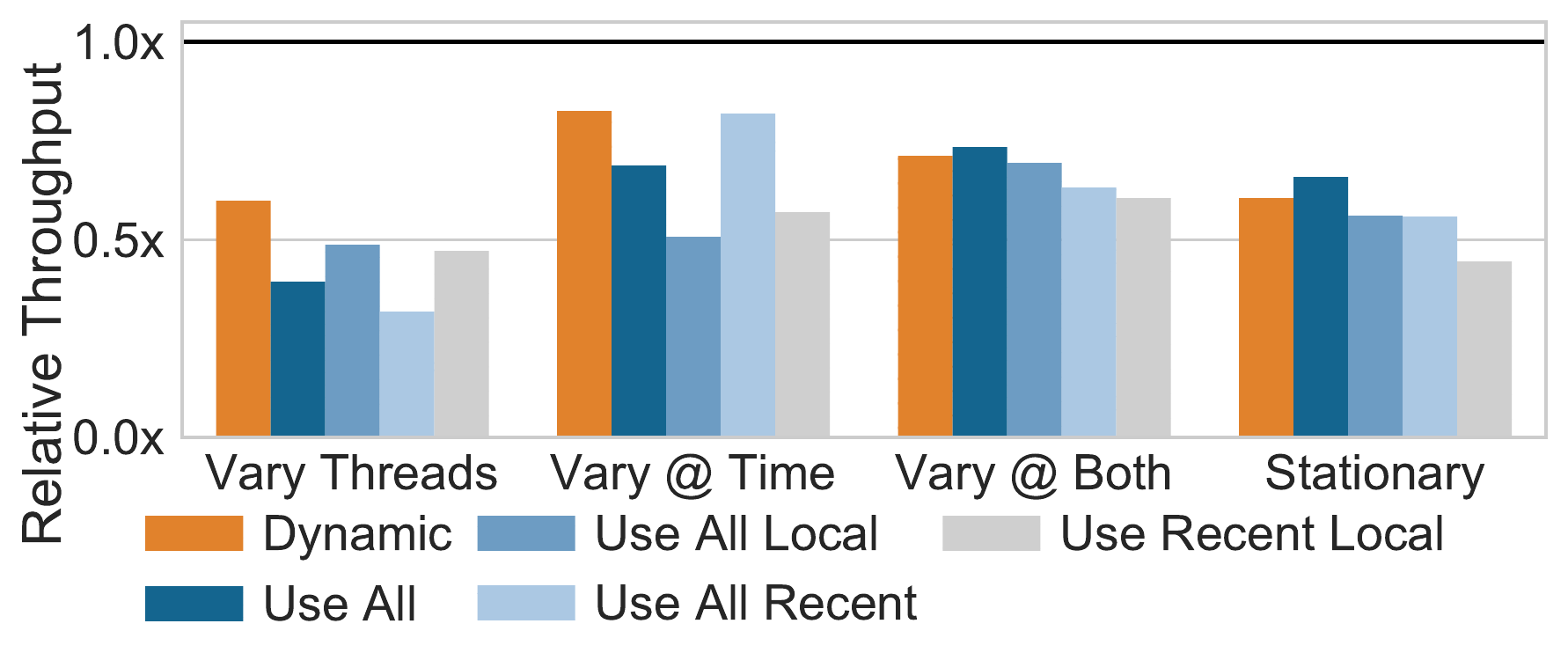}
\caption{Convolution Tuning in dynamically-changing workloads with context that partially captures the changes}
\label{fig:flickr_8node_nonstationary_somecontext}
\end{figure}

\begin{figure}[!t]
  \centering
  \includegraphics[width=1.0\columnwidth]{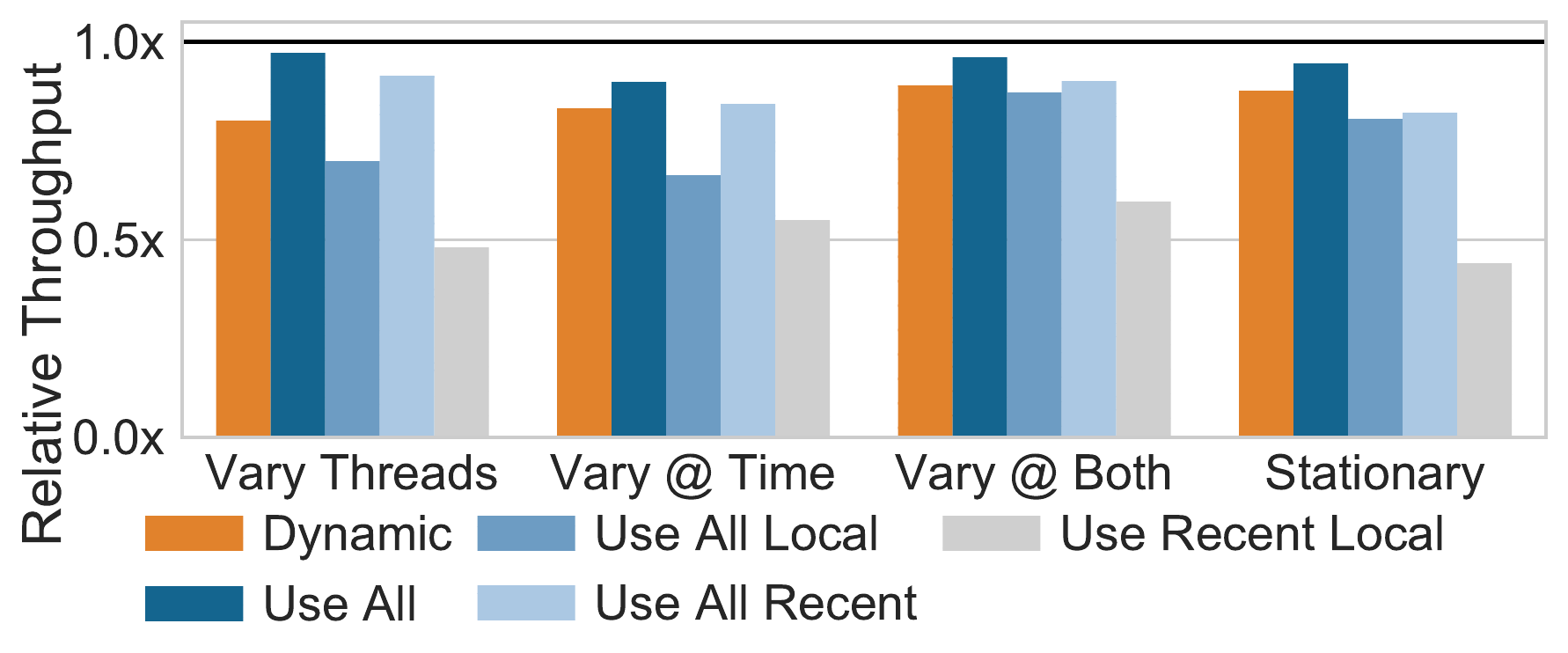}
\caption{Convolution Tuning in dynamically-changing workloads with context that fully captures the changes}
\label{fig:flickr_8node_nonstationary_allcontext}
\end{figure}

Using the same experimental setup as \ref{sec:dynamic-eval}, we test how well \cuttlefish{} can tune adaptive contextual convolution operators in dynamic settings where operator runtimes vary over time and across cores in the cluster. \Cref{fig:flickr_8node_nonstationary_somecontext} shows the relative throughput results when using a contextual adaptive convolution operator using some useful features, but not enough to fully model the dynamic changes in filter group dimensions that occur during the workloads. Importantly, the dynamic tuner manages to match or outperform the other strategies in the dynamic settings. However, this effect is less pronounced than for the context-free adaptive operator because some of the dynamic changes are being explicitly modeled. Interestingly, the large skew in convolution algorithm throughput across threads in the `Vary Threads' filter set cause this contextual operator to severely under-perform the context-free operator. This is because this contextual operator requires much more exploration to counteract it's insufficient set of contextual features. Finally, \Cref{fig:flickr_8node_nonstationary_allcontext} shows the relative throughput results when using a contextual convolution operator that captures enough features to fully model the underlying changes in the groups of filters used during the convolutions. This plot shows that when a contextual tuner is able to fully model all dynamic changes in the workload that affect the reward distributions, \cuttlefish{}'s default non-dynamic tuners should always be used.

\subsection*{D. System Overheads}

\begin{figure*}[!t]
 \centering
\begin{tabular}{lrr}
\toprule
Component &  Time & Space (Memory) \\
\midrule
Context-free tuner & $O(A)$ per tuning round & $O(A)$ bytes per \texttt{state} \\
Contextual tuner  & $O(A\cdot F^3)$ per tuning round  & $O(A\cdot F^2)$ bytes per \texttt{state} \\
Locking & $O(1)$ or $O(A)$ locks \& unlocks per tuning round & N/A \\
Context-free message      & $O(2\cdot N)$ messages per communication round & $O(A)$ bytes per message \\
Contextual message      & $O(2\cdot N)$ messages per communication round & $O(A\cdot F^2)$ bytes per message \\
State Aggregation at Model Store (context-free)   &   $O(A\cdot N^2)$ per communication round & $O(A\cdot N)$ bytes stored on Master \\
State Aggregation at Model Store (contextual)             &   $O(A\cdot F^2\cdot N^2)$ per communication round & $O(A\cdot F^2\cdot N)$ bytes stored on Master \\
\bottomrule
\end{tabular}
\caption{Naive asymptotic system overheads for \cuttlefish{} components}
\label{table:system_overheads}
\vspace{-0.15in}
\end{figure*}

To be useful across many real-world scenarios, it is important for \cuttlefish{} tuners to have low system overheads. We summarize the asymptotic time and space complexities of different components of \cuttlefish{} in \Cref{table:system_overheads}. $A$ is the number of choices a tuner is selecting between, $F$ is the feature count of a contextual tuner, and $N$ is the number of workers or agents in \cuttlefish{}'s distributed architecture. We set up a simple microbenchmark that tests the timing of unoptimized \cuttlefish{} tuner implementations using out-of-the-box Scala linear algebra libraries. We measured context-free tuners with five operator variants as taking $0.03$ milliseconds per choose-and-observe round. Contextual tuners with five operator variants and 2, 4, and 8 features respectively took $0.034$, $0.046$, and $0.082$ milliseconds per round. These results are in line with our observations from the real adaptive operators we tested. If these overheads are too high, multiple tuning rounds may be batched into a single round. However, a consequence of this is that tuning will require more rounds overall to be effective. 

The $O(N^2)$ term in the complexities of merging states during communication rounds occurs because $N\cdot (N-1)$ \texttt{state} merges happen during each communication round. Although this may appear to put scalability at stake, with five operator variants and ten contextual features, each individual \texttt{state} merge took well under 1 microsecond. This was the case even when applying the statistical test to handle dynamically-changing environments. So, this overhead is not immediately problematic. Additionally, in non-dynamically-changing environments, introducing a ``\texttt{state} un-merge'' method would allow doing the merging in $O(N)$, because only one global aggregation would need to be maintained, and sending a non-local \texttt{State} to each worker would only require un-merging the local worker state first. The overhead is most likely to become an issue on very large clusters when \cuttlefish{} is configured to handle dynamically-changing environments with one \cuttlefish{} agent per core in the cluster, and when storing a very large cache of multiple agent-epoch states per agent.

\end{document}